\pdfoutput=1
\documentclass[showpacs,amsmath,amssymb,aps,pre,twocolumn]{revtex4-1}
\usepackage{graphicx}% Include figure files
\usepackage{dcolumn}% Align table columns on decimal point
\usepackage{bm}% bold math
\usepackage[breaklinks,colorlinks = true,linkcolor = red,urlcolor=blue,citecolor=red]{hyperref}
\usepackage{multirow}
\usepackage{array}
\usepackage{booktabs}
\usepackage{ctable}
\usepackage{ulem}
\usepackage{upgreek}
\usepackage{epsfig}
\usepackage{mathrsfs}
\usepackage{amssymb}
\usepackage{amsbsy}
\usepackage{color}
\usepackage{cancel}
\usepackage{pifont}
\usepackage{marginnote}
\usepackage{float}
\usepackage{verbatim}
\usepackage{subfigure}
\usepackage{bbm, dsfont}
\usepackage{lineno}
\usepackage{amsmath}
\definecolor{dgreen}{rgb}{0,0.7,0}

\def\bea{\begin{eqnarray}}
\def\eea{\end{eqnarray}}

\def\nn{\nonumber}

\makeatletter

\newcommand{\Rmnum}[1]{\expandafter\@slowromancap\romannumeral #1@}
\makeatother

\newcommand{\eref}[1]{Eq.~(\ref{#1})}%
\newcommand{\fref}[1]{Fig.~\ref{#1}} %
\newcommand{\sref}[1]{Sec.~\ref{#1}}%
\newcommand{\aref}[1]{Appendix~\ref{#1}}%
\usepackage[utf8]{inputenc}

\begin{abstract}
While averages and typical fluctuations often play a major role to understand the behavior of a non-equilibrium system, this nonetheless is not always true. Rare events and large fluctuations are also pivotal when a thorough analysis of the system is being done. In this context, the statistics of extreme fluctuations in contrast to the average plays an important role, as has been discussed in fields ranging from statistical and mathematical physics to climate, finance and ecology. Herein, we study Extreme Value Statistics (EVS) of stochastic resetting systems  which have recently gained lot of interests due to its ubiquitous and enriching applications in physics, chemistry, queuing theory, search processes and computer science. We present a detailed analysis for the finite and large time  statistics of extremals (maximum and arg-maximum i.e., the time when the maximum is reached) of the spatial displacement in such system. In particular, we derive an exact renewal formula that relates the joint distribution of maximum and arg-maximum of the reset process to the statistical measures of the underlying process. Benchmarking our results for the maximum of a reset-trajectory that pertain to the Gumbel class for large sample size, we show that the arg-maximum density attains to a uniform distribution regardless of the underlying process at a large observation time. This emerges as a manifestation of the renewal property of the resetting mechanism. The results are augmented with a wide spectrum of Markov and non-Markov stochastic processes under resetting namely simple diffusion, diffusion with drift, Ornstein-Uhlenbeck process and random acceleration process in one dimension. Rigorous results are presented for the  first two set-ups while the latter two are supported with heuristic and numerical analysis. 
\end{abstract}

\begin{document}
%maketitle

\title{Extremal statistics for stochastic resetting systems}
\author{Prashant Singh$^{1}$ and Arnab Pal$^{2}$}
\email{prashant.singh@icts.res.in}
\email{arnabpal@mail.tau.ac.il}
\affiliation{\noindent \textit{$^{1}$ International Centre for Theoretical Sciences, Tata Institute of Fundamental Research, Bengaluru 560089, India}}

\affiliation{\noindent \textit{$^{2}$ School of Chemistry, The Center for Physics and Chemistry of Living Systems, Tel Aviv University, Tel Aviv 6997801, Israel}}
\date{\today}

\maketitle
\section{INTRODUCTION}
In many situations of physical relevance, extreme events are of tremendous importance despite they occur rarely. Starting from stock market crashes or large insurance losses in finance, records in Olympics to natural calamities such as earthquake, heat waves, extreme events or tsunamis -- all are typical examples of extreme events. Extreme Value Statistics (EVS) sits on the heart of a branch of statistics which deals with the probabilities generated by random processes responsible for such unusual extreme events \cite{EVS-review-1,EVS-review-2,EVS-review-3,EVS-review-4,EVS-review-5,EVS-review-6,EVS-review-7,EVS-review-cor}. The study of EVS has been extremely important in the field of disordered systems \cite{disorder-1,disorder-2}, fluctuating interfaces \cite{KPZ-1,KPZ-2}, interacting spin systems \cite{spin}, stochastic transport models \cite{EVS-con-1,EVS-STR}, random matrices \cite{RM-1,RM-2,RM-3}, ecology \cite{ecology}, in binary search trees \cite{tree} and related computer search algorithms \cite{CS-1,CS-2} and even in material science \cite{EVS-material}. We refer to these extensive reviews which provide detailed account of recent theoretical and application based progresses of EVS in science. In EVS, the famous Gnedenko’s classical law of extremes provides statistics of the maximum (or minimum) of a set of uncorrelated random variables (see e.g. \cite{EVS-review-1,EVS-review-2,EVS-review-3,EVS-review-4,EVS-review-5,EVS-review-6,EVS-review-7,EVS-review-cor}). However, there exists a myriad of systems for which the underlying random variables can be weakly or strongly correlated due to the correlations \cite{EVS-correlated-2,EVS-correlated-3,EVS-correlated-4} or a global conservation \cite{EVS-con-3,EVS-con-4,EVS-con-5,EVS-con-6} among the random variables, see \cite{EVS-review-cor} for a comprehensive review. One of the central goals of the subject is then to understand the statistics of extremes i.e., the maximum $M(t)$ of a given trajectory $x(t)$ which is observed upto time $t$ and the time $t_m$ to reach the maximum for such correlated systems namely Brownian motion and its generalizations \cite{Levy,Andersen,tmax-1,tmax-2,tmax-3,tmax-CTRW,tmax-RAP,tmax-CTRW}, run and tumble motion \cite{tmax-RTP-1,Mori2020}, fractional Brownian motion \cite{tmax-FBM-1,tmax-FBM-2,tmax-FBM-3}, random acceleration \cite{tmax-RAP}, anomalous walker \cite{tmax-anamolous} and fluctuating interfaces \cite{tmax-interface-growth}. Surprising enough, there are only a very limited exact results known on renewal processes (e.g., see \cite{EVS-con-3} where EVS for the longest waiting interval was analyzed and not the quantities of our interest). Moreover, prediction of limiting extreme value distributions in renewal processes has been extremely challenging (also see \cite{Restart1,review} and the discussion below). In this paper, we set out to understand in details the extremal statistics in a recently popularized renewal process namely stochastic resetting. 

Stochastic resetting is a renewal process in which the dynamics repeats by itself after random intervals controlled externally \cite{review}. The subject has recently gained considerable attention due to its vigorous applications in statistical physics \cite{Restart1,Restart2,Restart3,Restart4,Restart5,Restart6,PalJphysA,confining}, stochastic process \cite{SP-0,SP-1,SP-4,SP-5,SP-6} and other cross-disciplinary fields such as chemical and biological process \cite{ReuveniEnzyme1,bio-1,bio-2,bio-3}, computer science \cite{Luby,algorithm} and search theory \cite{HRS,Montanari,bressloff}. Brownian motion with resetting introduced by Evans and Majumdar in \cite{Restart1,Restart2} is the paradigmatic example of the subject which essentially captures two central features: emergence of a non-equilibrium steady state with a non-zero probability current \cite{Restart1,Restart2,Restart3,Restart4,Restart5,Restart6,PalJphysA} and expedition of a first passage time process \cite{PalReuveniPRL17,ReuveniPRL16,branching,Belan,Chechkin,interval,Peclet,space,interval-v}. Over the years, a large volume of work has been done to extend this simple model beyond diffusion to others such as underdamped \cite{underdamped} and scaled \cite{scaled} Brownian motion, random acceleration process \cite{RAP}, and active particles \cite{RTP-0,RTP-1,RTP-2}. Rigorous efforts have also been made to understand non-Poissonian strategies \cite{review,PalJphysA,PalReuveniPRL17,Chechkin}. The subject has also found interesting applications in stochastic thermodynamics \cite{sth-0,sth-1,sth-2}, quantum systems \cite{quantum-1}, many particle systems \cite{SEP,TASEP,Ising}, and nonlinear systems \cite{dynamical}. Very recently, resetting has also seen advances in single particle experiments using optical traps \cite{expt-1,expt-2}. We refer to this recent review \cite{review} and references therein for more details on the subject.

Notwithstanding that the subject has seen tremendous progress in statistical physics, exact results on EVS of stochastic resetting have been very limited. It has been understood that the maximum $M$ of a Brownian trajectory under stochastic resetting (i.e., Poissonian resetting) belongs to the Gumbel class for large sample size \cite{review,Restart1}. But only recently exact expressions for the first two moments of the maximum $M(t)$ have been obtained by Majumdar \textit{et al} in \cite{MajumdarMori2020}. Using these, the mean perimeter
and the mean area of a convex hull of the 2D Brownian motion with resetting were computed. On the other hand, when resetting times are taken from a power law density (i.e., non-exponential or non-Poissonian resetting), it was shown that the distribution of the maximum of the reset process is given by the Fr\'{e}chet law when appropriately centred and scaled \cite{Villarroel JStatMech}. A Weibull limit law was derived for first passage time under restart with branching \cite{branching}. But to the best of our knowledge, not much is known on the statistics for the arg-max i.e., the time $t_m$ to reach this maximum which is also a very important statistical measure (recall the famous arcsine law of L\'{e}vy in classical probability theory \cite{Levy}). Also exact results for maximum beyond simple diffusion are not available at this moment. This paper exactly aims to bridge this void. First, we derive a renewal formula (\ref{main-eq-5}) for the joint distribution of $M(t)$ and $t_m(t)$ in the presence of restart in terms of their underlying joint distribution and other statistical quantities. This formula is valid for any underlying stochastic process (Markov or non-Markov) as long as the memory is erased after each resetting event. Secondly, utilizing the renewal formula, we derive exact and asymptotic $n$-th order moments of $M(t)$ for diffusion and drift-diffusion process. We also show the convergence to the Gumbel limit law for both these cases. Third and importantly, we obtain exact expression for the moment generating function for $t_m$ from which we show that the density of $t_m$ pertains to a \textit{universal} $1/t$ form \textit{independent of the underlying process} at large time with sub-leading process dependent corrections. We first demonstrate this result exactly for diffusion and drift-diffusion process and then generalize to arbitrary stochastic process.

The remainder of the paper is structured as follows. We derive the joint distribution of $M(t)$ and $t_m(t)$ in the Laplace space of $t~(\to s)$ and $t_m~(\to k)$ in \sref{prelim}. This renewal formula becomes instrumental to investigate the statistics of $M(t)$ for simple diffusion and drift-diffusion process in \sref{statistics-of-M}. In particular, we compute the exact moments for the maximum $M(t)$ and then present the asymptotic limiting distributions. In \sref{statistics-of-argmax}, we present our results for arg-max $t_m(t)$ for simple diffusion and drift-diffusion process. We compute the moments and analyze the large time behavior for the arg-max density. The large time limiting density is shown to be universal which is further proven in \sref{gen-process} for generic stochastic process augmented with numerical simulations. We summarize our results in \sref{conclusion}. For brevity, many supplemented derivations of our results have been reserved to the Appendix.

\section{Renewal formula for JOINT DISTRIBUTION OF $M$ and $t_m$}
\label{prelim}
\begin{figure}[t]
\includegraphics[scale=0.42]{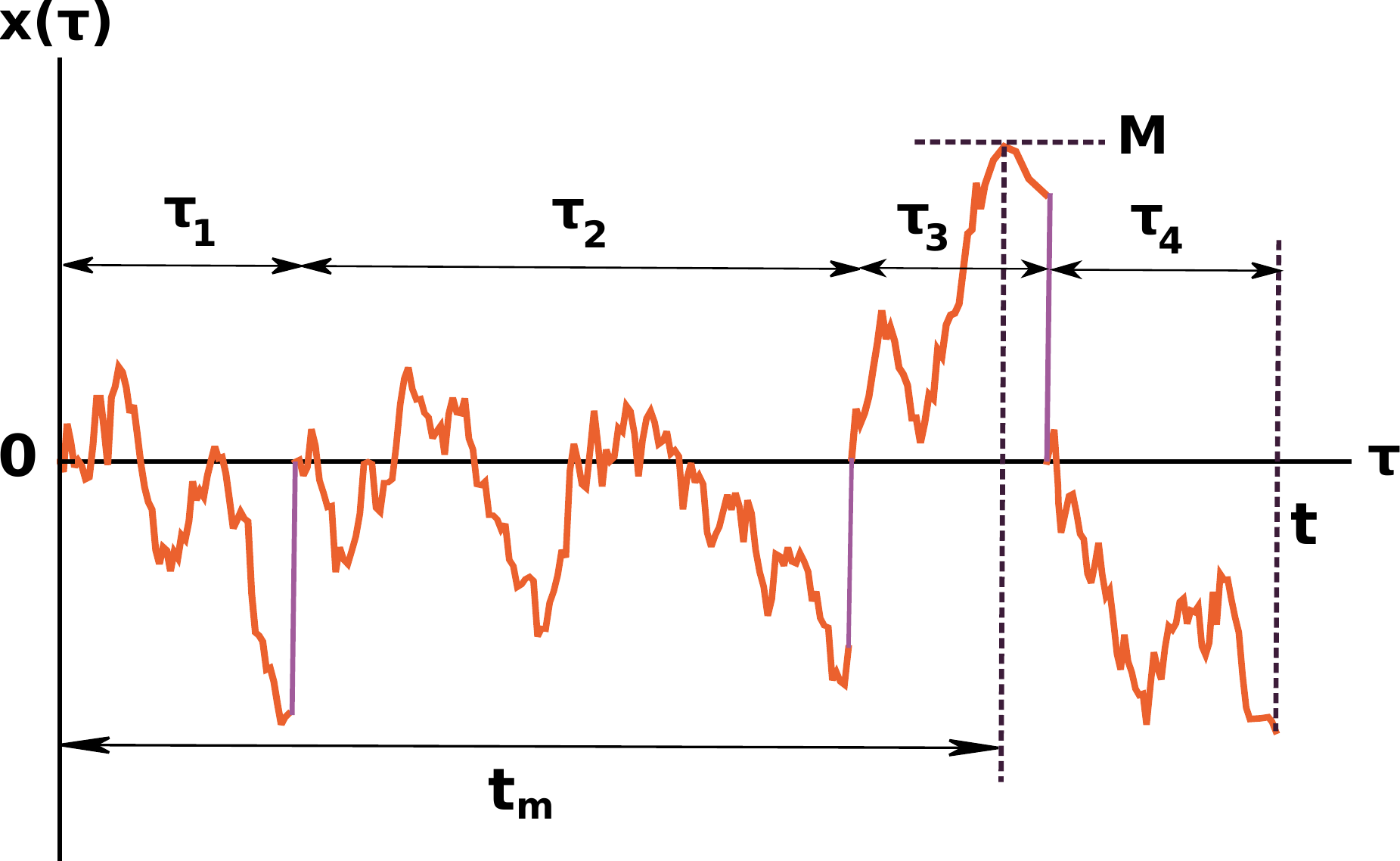}
\centering
\caption{Schematic of the maximum distance $M$ and the time $t_m$ to reach $M$ for a trajectory  $x(\tau)$ of a given stochastic process observed upto time $t$. The vertical solid lines indicate the resetting events which take place at times $\tau _1,~\tau_2,\cdots$ taken from a distribution $p(\tau)$. The resetting coordinate is same as the initial condition which is set to be the origin here. Total number of resetting intervals in this realization is $4$, and the maximum occurs during the third interval.}
\label{trajectory-pic}
\end{figure}

We begin with the derivation of the joint distribution for $M$ and $t_m$ in the presence of resetting. Consider a typical trajectory $x(\tau)$ of a particle governed by some stochastic law of motion and observed upto a fixed time $t$ (see \fref{trajectory-pic}). Motion of the particle is also subjected to resetting that brings it back to the origin at a constant rate $r$. This essentially means that the waiting time between any two resetting events is taken from the distribution $p(\tau)=re^{-r\tau}$. Note that, in a fixed time window $t$, the number of resetting undergone by the particle is a random variable and varies from trajectory to trajectory. Let us assume that the trajectory is divided into $N$-intervals out of which the particle has experienced $N-1$ resetting events and further denote the waiting times in these intervals as $\tau_i$ with $i=1,2,...,N-1$. However, in the last intervals $\tau_N$, the particle does not experience any resetting event: the probability of which is given by $\int_{\tau_N}^\infty~p(\tau)d\tau=e^{-r\tau_N}$. Since the observation time is fixed, $N$ is a random variable and varies from trajectory to trajectory. Also, without any loss of generality, we assume the starting point as a resetting event which gives $N \geq 1$. 

We now proceed to compute the joint distribution of $M,~t_m$ and $N$ which is denoted by $\mathcal{P}_r \left(M,t_m,N|t\right)$. Note that the maximum distance $M$ can be attained during any of the $N$ time intervals $\{\tau _i \} = \{\tau _1, \tau_2,...,\tau_N \}$. For example, if the maximum $M$ is attained at time $\tau$ during the the first interval then the contribution to $\mathcal{P}_r \left(M,t_m,N|t\right)$ is given by
\begin{align}
\mathcal{C}_1=    &\left(\prod _{i=1}^{N}\int_0^\infty d\tau_i\right)~~\int_0^{\infty}~d\tau~P_0(M,\tau|\tau_1) p(\tau_1) \nonumber \\ 
    &\times \left[\prod_{i = 2}^{N-1}~S_0(0,\tau_i|M) p(\tau_i) \right] \times \left[ e^{-r\tau_N} S_0(0,\tau_N|M) \right] \nonumber \\
    &~~~~~~~~~~\times  \delta(\tau-t_m)~\delta(t-\sum_{i=1}^N \tau_i),
    \label{contr-1}
\end{align}
where we introduce $P_0(M,t_{m}|t)$ as the joint distribution of $M$ and $t_{m}$ for the underlying process (without resetting) till time $t$. Moreover, $S_0(x_0,\tau _j|M)$ is the probability that the particle starting from $x_0$ always stays below $M$ upto time $\tau_j$ in the absence of resetting. In other words, this is the survival probability that a particle, starting from $x_0$, survives an absorbing boundary at $M(>x_0)$ till time $\tau_j$. Rationale behind various terms in $\mathcal{C}_1$ can be understood in the following way: the maximum $M$ is attained by the underlying process in the first interval that lasts for time $\tau_1$ which gives rise to $P_0(M,\tau|\tau_1) p(\tau_1)$ term in the first line of Eq. \eqref{contr-1}. Essentially, in the remaining $N-1$ intervals, the trajectory always stays below $M$. As a consequence, we get $N-1$ survival probabilities appropriately weighted in time in the second line. Finally, the first $\delta$-function asserts the condition $t_m=\tau$ which is true here while the second one ensures that the total observation time is $t$.

Similarly, we can write the contribution to $\mathcal{P}_r \left(M,t_m,N|t\right)$ when the maximum $M$ is attained in the second interval at time $\tau$ while the particle remains below $x=M$ during the other intervals. This contribution is given by 
\begin{align}
  \mathcal{C}_2 = &  \left(\prod _{i=1}^{N}\int_0^\infty d\tau_i\right)~\int_0^{\infty}~d\tau~P_0(M,\tau|\tau_2) p(\tau_2)\nonumber\\
  &\times \left[ \prod_{i =1, i \neq 2}^{N-1}~S_0(0,\tau_i|M) p(\tau_i)\right] \times \left[ e^{-r\tau_N} S_0(0,\tau_N|M) \right]\nonumber \\ &~~~~~~~ \times \delta(t-\sum_{i=1}^N \tau_i)~ \delta(\tau_1+\tau-t_m).
  \label{contr-2-SM}
\end{align}
Following the same physical argument, one can also write the contributions $\mathcal{C}_3,\mathcal{C}_4..$ for the maximum to be in the second, third,... interval respectively. In particular, when the maximum is in the last reset-free interval $\tau_N$, we have
\begin{align}
  \mathcal{C}_N= & \left(\prod _{i=1}^{N}\int_0^\infty d\tau_i\right)~\int_0^{\infty}~d\tau~P_0(M,\tau|\tau_N) e^{-r\tau_N} \nonumber \\
  &~~~~~~~~~~ \times \left[ \prod_{i =1}^{N-1}~S_0(0,\tau_i|M) p(\tau_i) \right]~ \nonumber \\
  & ~~~~~~~~\times \delta(t-\sum_{i=1}^N \tau_i)~ \delta(\sum_{i=1}^{N-1}\tau_i+\tau-t_m).
\label{contr-3}
\end{align}
Thus, the joint distribution $\mathcal{P}_r \left(M,t_m,N|t\right)$, can be obtained by summing over all the contributions $\mathcal{C}_1,~\mathcal{C}_2,...,\mathcal{C}_N$. Performing the sum, $\mathcal{P}_r \left(M,t_m,N|t\right)$ can be formally written as
\begin{align}
\mathcal{P}_r\left(M, t_{m},N|t \right) &= \frac{1}{r}\sum _{j=1}^{N} \int_{0}^{\infty} d \tau _j~  d \tau ~P_0\left(M, \tau|\tau _j \right) p(\tau _j) \nonumber \\
&\times \left( \prod _{j'=1,j' \neq j}^{N}\int_{0}^{\infty} d \tau _{j'} S_0(0, \tau _{j'}|M) p(\tau _{j'}) d \tau _{j'}\right) \nonumber \\ 
& \times \delta \left( \sum _{i=1}^{j-1}\tau _i+\tau - t_{m} \right)\delta(t-\sum_{i=1}^N \tau_i).
\label{main-eq-2}
\end{align}

To proceed further, it is only natural to take the Laplace transformations with respect to $t~(\to s)$ and $t_m~(\to k)$. Denoting the Laplace transform of $\mathcal{P}_r\left(M, t_{m},N|t \right)$ by $\mathcal{Z}_r\left(M, k, N|s \right)$ and further performing the sum over all the intervals $N$, we find (see \aref{Joint-LT})
{\footnotesize
\begin{align}
\mathcal{Z}_r\left(M, k|s \right)
&= \frac{\mathcal{Z}_0\left(M, k|r+s \right)}{\left[1-r \bar{S}_0(0, s+r|M)\right]~\left[1-r\bar{S}_0(0, s+r+k|M) \right]},
\label{main-eq-5}
\end{align}}
where $\mathcal{Z}_r\left(M, k|s \right)$ is the Laplace transform of the joint distribution $P_r(M,t_{m}|t)$ (subscript $0$ will indicate the same without resetting) and $\bar{S}_0(0, s|M)=\int_0^\infty~dt~e^{-st}S_0(0, t|M)$ is the Laplace transformation of the underlying survival probability. \eref{main-eq-5} is the first central result of our paper. Such a renewal formula is very important since it relates the joint distribution of $M$ and $t_m$ with resetting to the underlying joint distribution and the survival probabilities. For Markov processes, one can also use the path-decomposition method \cite{tmax-1,tmax-2,tmax-3,MajumdarMori2020} to arrive at \eref{main-eq-5}. However, our derivation is more robust since it holds even when underlying process is non-Markov while the path-decomposition method strictly relies on the Markov property as was illustrated for Brownian motion by Majumdar \textit{et al} in \cite{MajumdarMori2020}. The only assumption that goes into the derivation is that the process does not retain memory between the resetting intervals. Finally, we remark that although \eref{main-eq-5} has been derived under a one-dimensional framework, it is valid also in higher dimensions. Also see \cite{Mori2020} where a similar approach has been used to study the persistent properties of run and tumble particles in arbitrary dimensions.

In what follows, we use the renewal formula in Eq. \eqref{main-eq-5} to study statistics of $M$ and $t_{m}$ for simple diffusion and drift-diffusion process in one dimension, respectively described by
\begin{align}
 \frac{dx}{d\tau}&=\eta(\tau)~\label{SD},\\
    \frac{dx}{d\tau}&=v+\eta(\tau)~\label{DDiff},
\end{align}
where $\eta(\tau)$ is the Gaussian white noise with mean $\langle \eta (\tau) \rangle =0$ and correlation $\langle \eta (\tau) \eta (\tau') \rangle =2D\delta (\tau-\tau')$, and further assume the drift-velocity $v>0$. Here, $D$ is the diffusion constant (which will be set to $\frac{1}{2}$ without any loss of generality for the rest of the paper). We also consider that the particle starts from the origin at $t=0$ and is reset to the origin at random times drawn from the distribution $p(\tau)=r e^{-r \tau}$.

\begin{figure*}[t]
  \centering
  \subfigure{\includegraphics[scale=0.22]{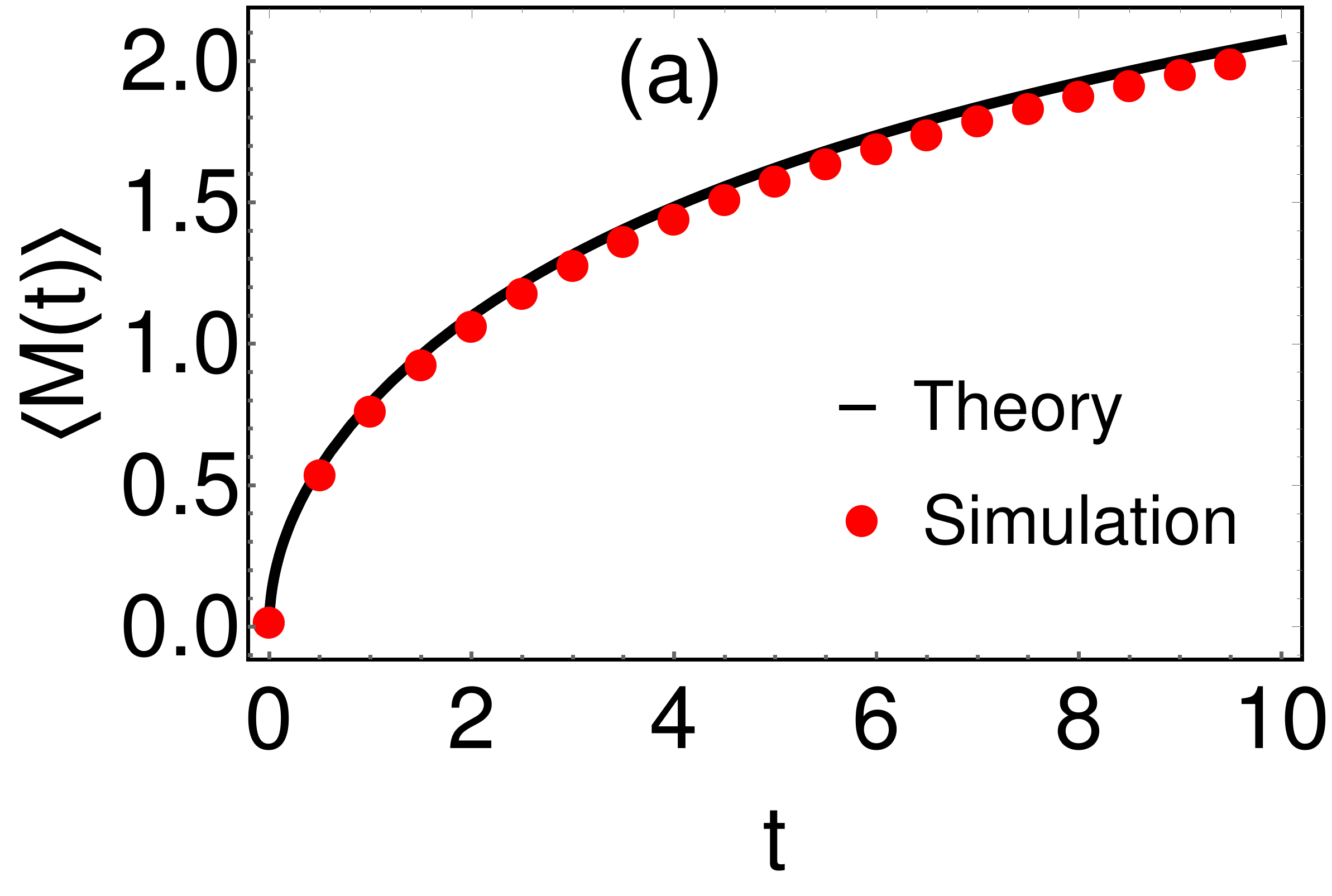}}
  \subfigure{\includegraphics[scale=0.22]{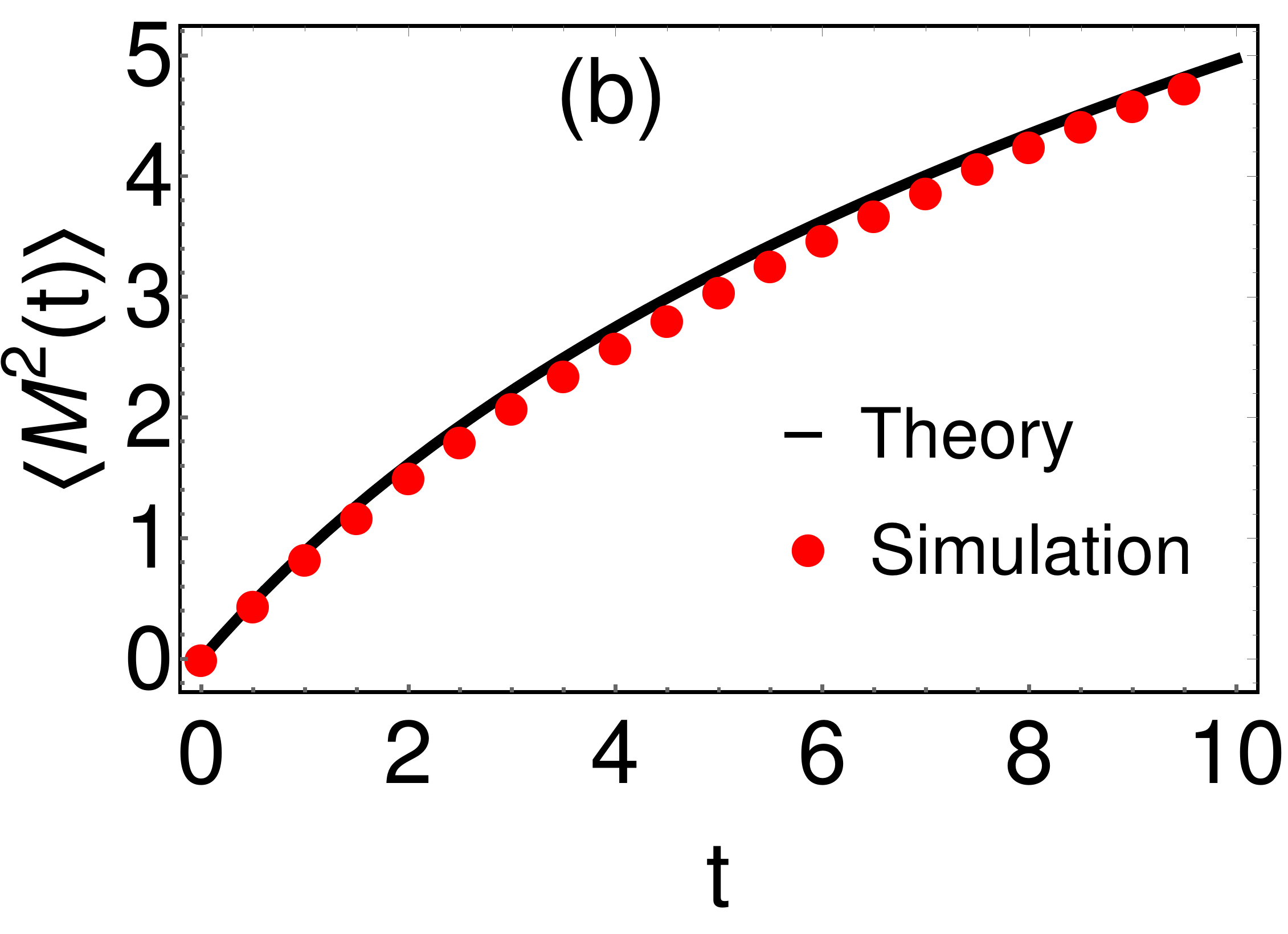}}
  \subfigure{\includegraphics[scale=0.22]{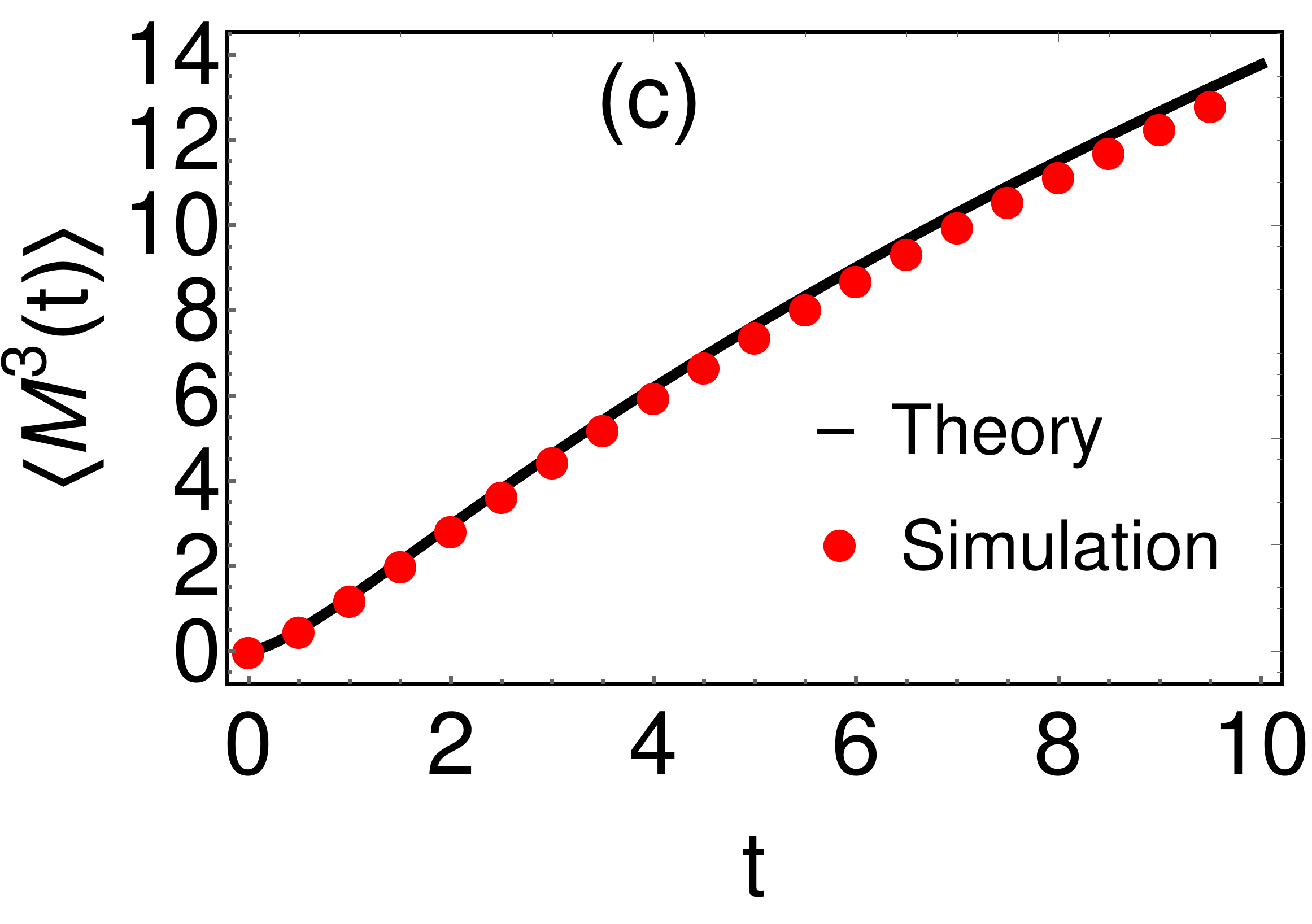}}
\centering
\caption{Comparison of moments for the maximum $\langle M^{n}(t) \rangle$ in Eq. \eqref{BMeq-6} for the simple diffusion against the numerical simulations for (a) $n=1,$ (b) $n=2$ and (c) $n=3$. We have fixed $r=1$ for all the simulations.}    
\label{BM-max-mom} 
\end{figure*}

\section{Statistics of maximum $M$}
\label{statistics-of-M}
In this section, we use Eq.\eqref{main-eq-5} to analyse the statistical properties of $M$ for the above-mentioned stochastic processes. To compute the distribution of $M$, one needs to integrate $P_r \left(M, t_m|t \right)$ over all $t_m$. This is equivalent to putting $k=0$ in $\mathcal{Z}_r\left(M, k|s \right)$ in Eq. \eqref{main-eq-5}. Let us consider the case of simple diffusion first.

\subsection{Simple diffusion}
\label{BM}
It is instructive to first review some known results on $M$ and $t_m$ without resetting which will be useful for subsequent studies. For simple diffusion, the joint distribution of $M$ and $t_m$ is given by \cite{tmax-CTRW}
\begin{align}
P_0\left( M,t_m|t\right) = \frac{M }{\pi t_m^{3/2}\sqrt{t-t_m}}e^{-\frac{M^2}{2 t_m}}.
\label{joint-BM}
\end{align}
Integrating over $M$, we get the distribution of $t_m$ as $P_0(t_m|t) = \frac{1}{\pi \sqrt{t_m(t-t_m)}}$ or equivalently the cumulative distribution
\begin{align}
\text{Prob}[t_m \leq T]=\int_0^T dt_m'~P_0(t_m'|t)=\frac{2}{\pi} \sin^{-1}\left[ \sqrt{\frac{T}{t}}  \right],
\label{arc-sine}
\end{align}
which is the celebrated `Arc-sine' law for the Brownian motion due to L\'{e}vy \cite{Levy}. To use Eq.\eqref{main-eq-5}, we need to specify the Laplace transformations $\bar{S}_0(0, s|M)$ and $\mathcal{Z}_0\left(M, k|s \right)$. The latter can be computed by taking the Laplace transformation of Eq. \eqref{joint-BM} with respect to $t$ and $t_m$ and this gives
\begin{align}
\mathcal{Z}_0\left(M, k|s \right) = \sqrt{\frac{2}{s}} e^{-\sqrt{2(s+k)}M} \label{BMeq-1}.
\end{align}
On the other hand, survival probability for a Brownian particle is a canonical result due to L\'{e}vy which reads $S_0(0,t|M)=\text{Erf}\left(\frac{M}{\sqrt{4Dt}} \right)$, where $\text{Erf}(z)=\frac{2}{\sqrt{\pi}}\int_0^z~e^{-y^2}dy$ is the error-function. Taking the Laplace transform, one gets
\begin{align}
\bar{S}_0(0, s|M) = \frac{1}{s} \left( 1-e^{-\sqrt{2s} M}\right) \label{BMeq-2}.
\end{align}
Inserting Eqs. \eqref{BMeq-1} and  \eqref{BMeq-2} in the renewal Eq. \eqref{main-eq-5} yields the Laplace transformation $\mathcal{Z}_r\left(M, k|s \right)$ from which one gets
\begin{align}
\mathcal{Z}_r\left(M, k=0|s \right) = \frac{\sqrt{2} \left( r+s\right)^{3/2} ~e^{-\sqrt{2(r+s)}M}}{ \left( s+r e^{-\sqrt{2(r+s)}M}\right)^2}.
\label{dist-max-BM-eq-1}
\end{align}
Note that this was also obtained in \cite{MajumdarMori2020} using the path decomposition method. To get the distribution of $M$ in the time domain, one has to perform the inverse Laplace transform of $\mathcal{Z}_r\left(M, k=0|s \right)$ with respect to $s$. Before that, we look at the moments of $M$ to get the effect of resetting on $M$. The $n$-th moment can be written in terms of $\mathcal{Z}_r\left(M, k=0|s \right)$ as 
\begin{align}
\int _0 ^{\infty} dt~ e^{-st} \langle M^{n}(t) \rangle& = \int _{0}^{\infty} dM \mathcal{Z}_r\left(M, k=0|s \right),\nonumber \\
& = -\frac{n!}{2^{\frac{n}{2}}r (r+s)^{\frac{n-2}{2}}}~ \text{Li}_n \left(-\frac{r}{s} \right), \nonumber \\
&=-\frac{n!}{2^{\frac{n}{2}}r (r+s)^{\frac{n-2}{2}}}~\sum _{k=1}^{\infty} \frac{(-r)^n}{k^n ~s^{k+1}},
\label{BMeq-5}
\end{align}
where $\text{Li}_n \left(-\frac{r}{s} \right)$ is the PolyLog function in the second line \cite{Table} and while going to the third line, we have used the series representation $\text{Li}_n \left(-y \right) = \sum _{k=1}^{\infty} \frac{(-y)^k}{k^n}$. Next, to get the moments in the time domain, we use the following inverse Laplace transformation:
\begin{align}
\mathcal{L}_{s \to t}^{-1} \left[ \frac{1}{(r+s)^{\frac{n-2}{2}}~s^{k+1}}\right] = \frac{_1\bar{F}_1 \left(-1+\frac{n}{2}, k+\frac{n}{2}, -r t \right)}{t^{1-k -\frac{n}{2}}},
\label{laplace-eq-1}
\end{align}
where $_1\bar{F}_1(a,b,z)$ stands for the regularized hypergeometric function \cite{Table}. Inserting this in Eq. \eqref{BMeq-5}, we find that $\langle M^{n}(t) \rangle$ possesses the scaling form
\begin{align}
\langle M^{n}(t) \rangle = \frac{1}{(\sqrt{2 r})^n} H_n (r t),
\label{BMeq-6}
\end{align}
with the scaling function $H_n(z)$ given by
\begin{align}
H_n(z) = (-1)^{-\frac{n}{2}} n! \sum _{k=1}^{\infty} \frac{~_1\bar{F}_1 \left(-1+\frac{n}{2}, k+\frac{n}{2}, -z \right)}{k^n(-z)^{1-k-\frac{n}{2}}}.
\label{BMeq-7}
\end{align}
Note again that the first two moments $(n=1,2)$ were recently obtained in \cite{MajumdarMori2020}. Our results are consistent with that in \cite{MajumdarMori2020} and further extend to obtain exact expressions for all moments of $M$. In Figure \ref{BM-max-mom}, we have plotted $\langle M^{n}(t) \rangle$ for $n=1,2,3$ and compared them against the numerical simulation. We observe excellent match between them. To illustrate the effect of resetting, it is instructive to look at the asymptotic behaviours of the scaling function $H_n(z)$ for which we analyse $\mathcal{Z}_r\left(M, k=0|s \right) $ in Eq. \eqref{dist-max-BM-eq-1} in various limits of $s r^{-1}$ and then appropriately use Eq. \eqref{BMeq-5} (first line) to obtain the moments of $M$. For the continuity of the presentation, we have relegated this derivation to appendix \ref{asy-sca-M-BM} and present only the final results here. The asymptotic forms read   
\begin{align}
H_n(z) &\simeq  \log ^n z+O \left( \log ^{n-1} z \right) , ~~~~~~~~\text{as } z \to \infty, \label{BMeq-8}\\
& \simeq \frac{n!}{\Gamma \left(1+\frac{n}{2} \right)} z^{\frac{n}{2}}+\mathcal{B}_n z^{\frac{n}{2}+1},~~~~~\text{as } z \to 0, \label{BMeq-9}
\end{align}
where the exact expression of $\mathcal{B}_n$ is given in Eq. \eqref{Bn-exp}. Finally, we insert Eqs. \eqref{BMeq-8} and \eqref{BMeq-9} in Eq. \eqref{BMeq-6} to get the behaviour of $\langle M^{n}(t) \rangle$ at large and short times as
\begin{align}
\langle M^{n}(t) \rangle &\simeq \frac{ \log ^n r t}{(\sqrt{2r})^n}  +O \left[ \log ^{n-1} r t \right], ~~~~\text{as } t\gg \frac{1}{r}, \label{BMeq-10}\\
& \simeq \frac{n! ~t^{\frac{n}{2}}}{2 ^{\frac{n}{2}}\Gamma \left(1+\frac{n}{2} \right)}+\frac{\mathcal{B}_n r t^{\frac{n}{2}+1}}{2^{n/2}} ,~~~\text{as } t  \ll \frac{1}{r}.
\label{BMeq-11}
\end{align}
For $r=0$, the moments in Eq.\eqref{BMeq-11} match, as expected, with that of the free Brownian motion \cite{tmax-CTRW}. It is worth noting that there is a crossover in $\langle M^{n}(t) \rangle$ from $ \sim t^{n/2}$ behaviour to $\sim \log ^{n} rt$ behaviour at time scale $t \sim \frac{1}{r}$ with the crossover function given exactly in Eq. \eqref{BMeq-7}. Although a diffusing particle approaches a non-equilibrium steady state in the presence of resetting \cite{Restart1}, the maximum $M(t)$ still increases with time but rather slowly (with logarithmic growth) for $t \gg \frac{1}{r}$. This same observation was also made recently by Majumdar \textit{et al} in \cite{MajumdarMori2020}. The logarithmic growth of the maximum in presence of resetting can also be understood heuristically from the extreme value statistics of weakly correlated variables which we illustrate later.

\begin{figure}[t]
\centering
  \includegraphics[scale=0.3]{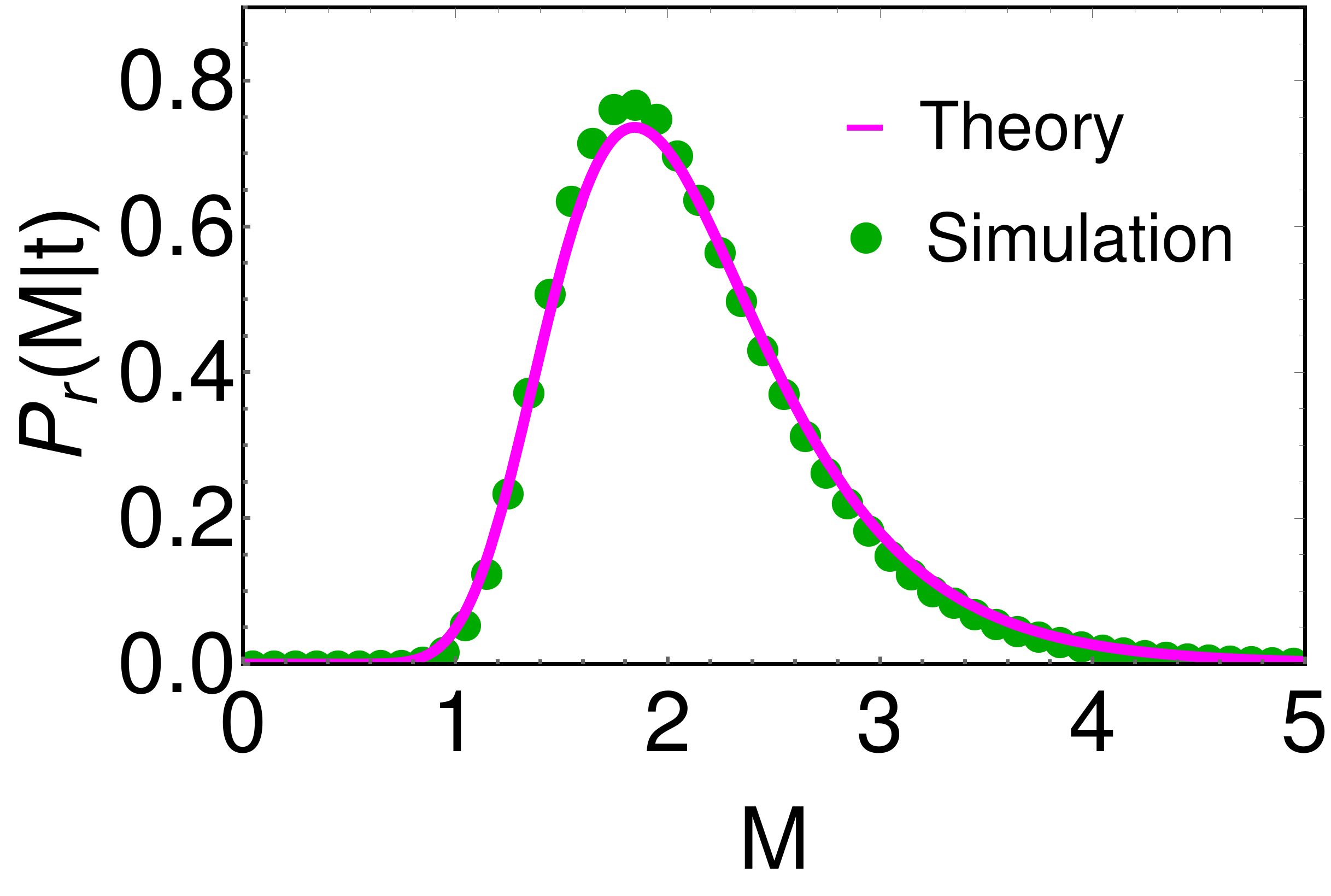}
  \includegraphics[scale=0.3]{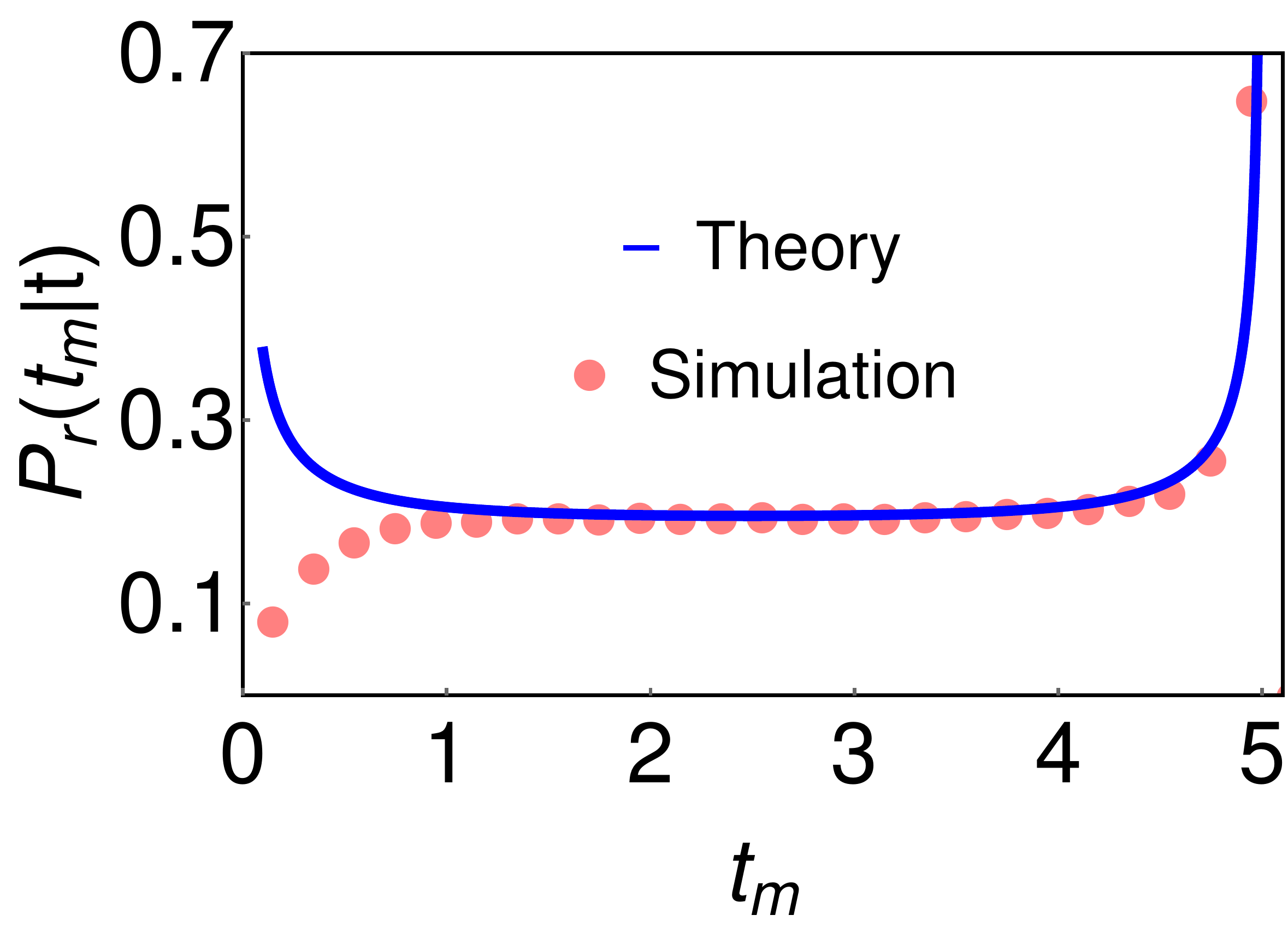}
  \caption{\textit{Top panel:} plot for the distribution of maximum $P_r(M|t)$ in Eq. \eqref{dist-M-BM} (in solid line) for the simple diffusion against the numerical simulations (in marker). For this plot, we fix: $r=2$ and $t=20$. \textit{Bottom panel:} comparison between the asymptotic form of  $P_r (t_{m}|t)$ in Eq. \eqref{BMeq-15-new} for the simple diffusion with with the numerical distribution. Since the theory works only in the large $t_m$ limit, we see a deviation when $t_m$ is small. Parameters set for this plot are: $r =1,~t=5$.}
  \label{BM-dist}
\end{figure}

After looking at the moments, we now consider the distribution of $M$ for which we have to invert $\mathcal{Z}_r\left(M, k=0|s \right)$ in Eq. \eqref{dist-max-BM-eq-1} with respect to $s$. Performing inversion for arbitrary $t$ turns out to be challenging. However for large $t$ (or equivalently small $s$), we can make some analytic progress. For $s \ll r$, we approximate $r+s \simeq r$ and use this in Eq. \eqref{dist-max-BM-eq-1} to get
 \begin{align}
\mathcal{Z}_r\left(M, k=0|s \right) \simeq  \frac{\sqrt{2} r^{3/2} ~e^{-\sqrt{2r}M}}{ \left( s+r e^{-\sqrt{2 r}M}\right)^2}.
\label{BMeq-11_new}
\end{align}
To get the distribution in the time domain, we use the inverse Laplace transform $\mathcal{L}_{s \to t} ^{-1} \left[ \frac{1}{(s+a)^2}\right] = t e^{-a t}$ for $a \geq 0$ in Eq. \eqref{BMeq-11_new} which yields
\begin{align}
P_r(M|t) \simeq \sqrt{2r^3} ~t e^{-\sqrt{2r}M} \text{exp} \left(- r t e^{-\sqrt{2r}M}  \right).
\label{dist-M-BM} 
\end{align}
We emphasize that the approximate equality in Eq. \eqref{dist-M-BM} indicates that this equation is valid only for $t \gg \frac{1}{r}$ when the effect of resetting is highest. In Fig. \ref{BM-dist} (top panel), we have compared $P_r(M|t)$ in Eq. \eqref{dist-M-BM} with the same obtained from the numerical simulations. We observe an excellent agreement between them. To understand Eq. \eqref{dist-M-BM} heuristically, we remark that the survival probability $S_r (0,t|M)$ for Brownian motion under reset possesses the Gumbel form at large times. Based on the extreme value statistics of weakly correlated variables, it was shown that $S_r (0,t|M) \simeq e^{- r t e^{-\sqrt{2 r} M}}$ for $t \gg \frac{1}{r}$ \cite{Restart1}. Since $S_r (0,t|M)$ is the cumulative distribution of $M$, one can appropriately differentiate it with respect to $M$ to get the distribution $P_r(M|t)$ in Eq. \eqref{dist-M-BM}. Furthermore, one can easily check that that using $P_r(M|t)$ from Eq. \eqref{dist-M-BM}, we get the same form of moments at large time as given in Eq. \eqref{BMeq-10}.

\begin{figure*}[t]
  \centering
  \subfigure{\includegraphics[scale=0.22]{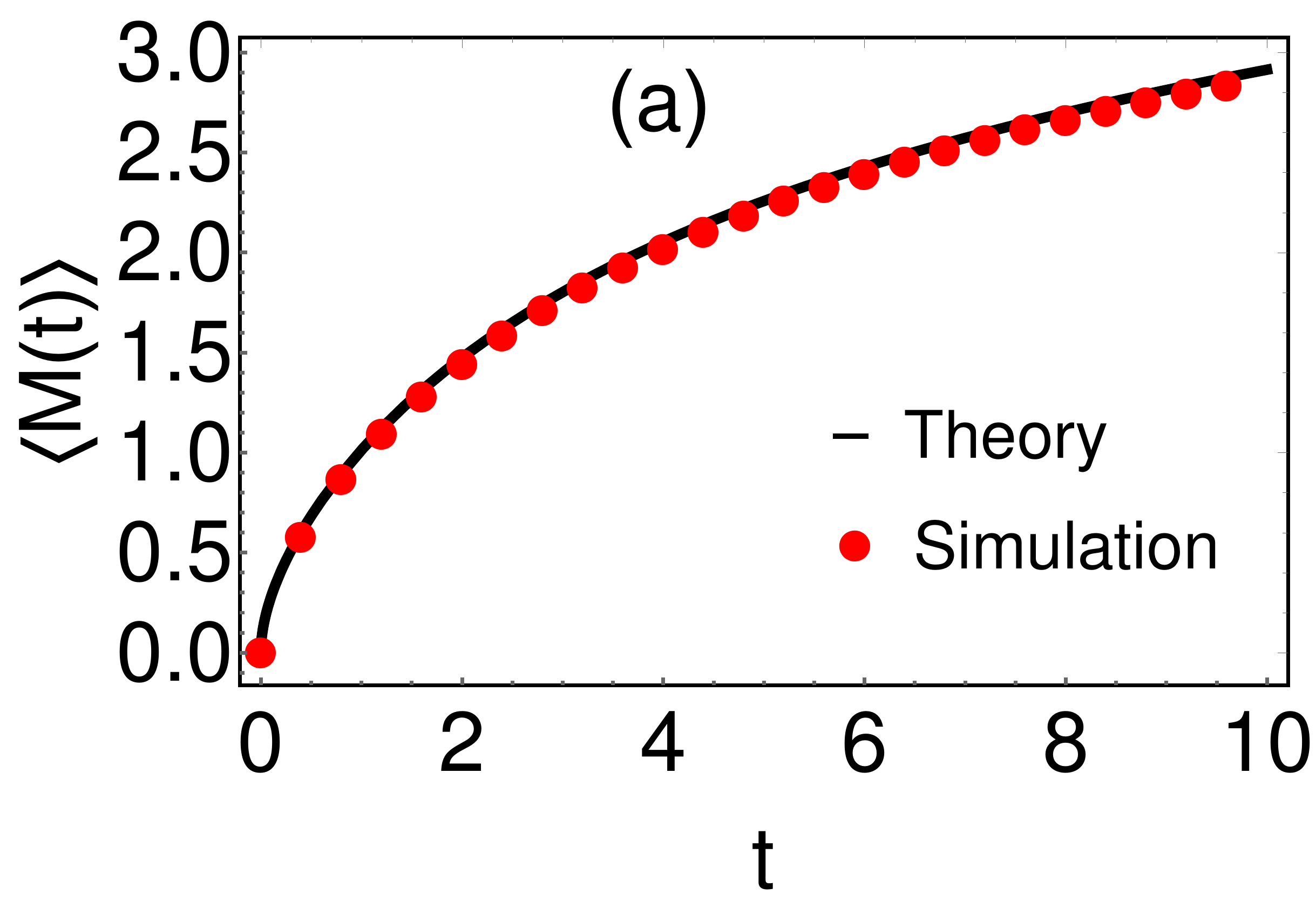}}
  \subfigure{\includegraphics[scale=0.22]{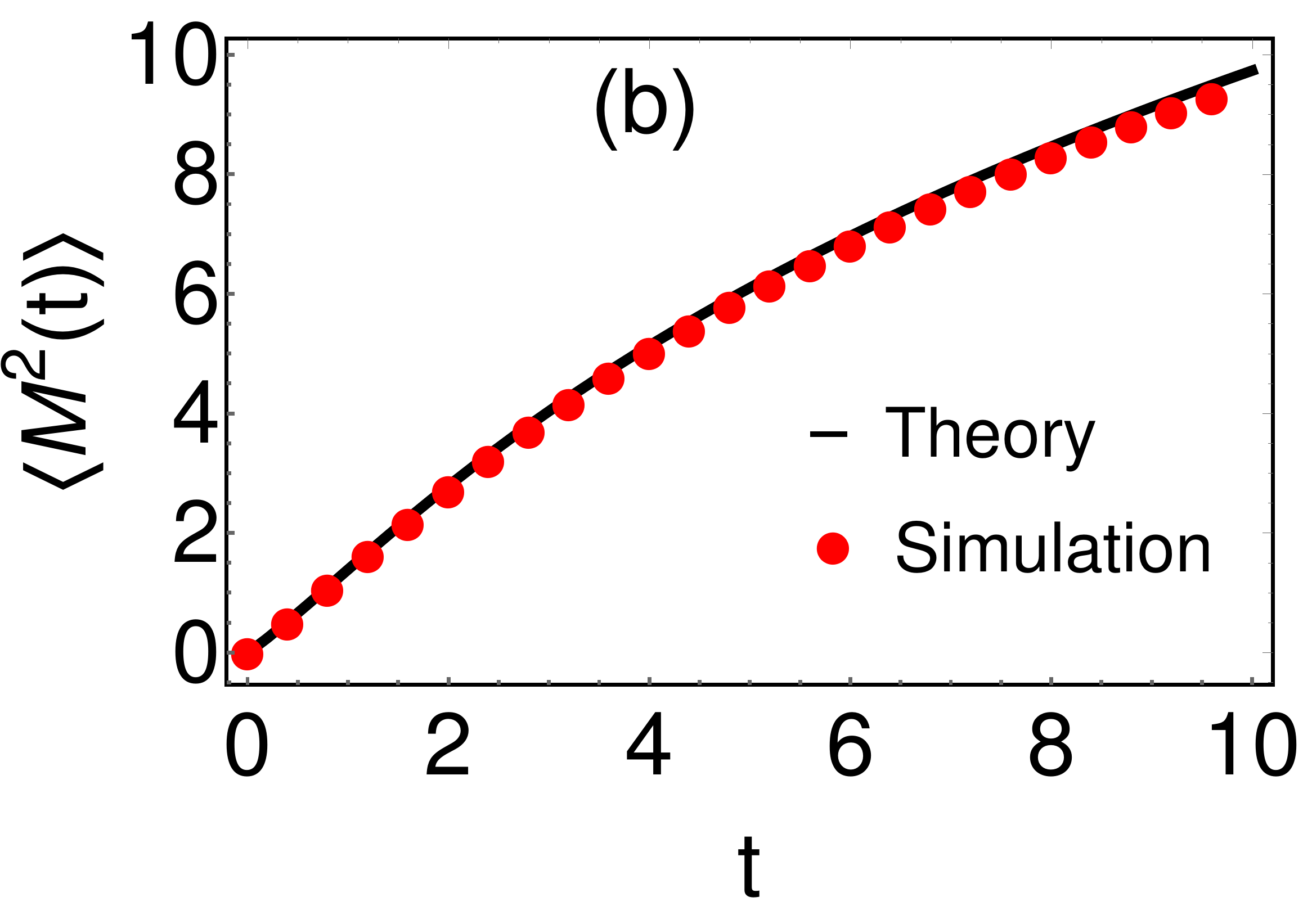}}
  \subfigure{\includegraphics[scale=0.22]{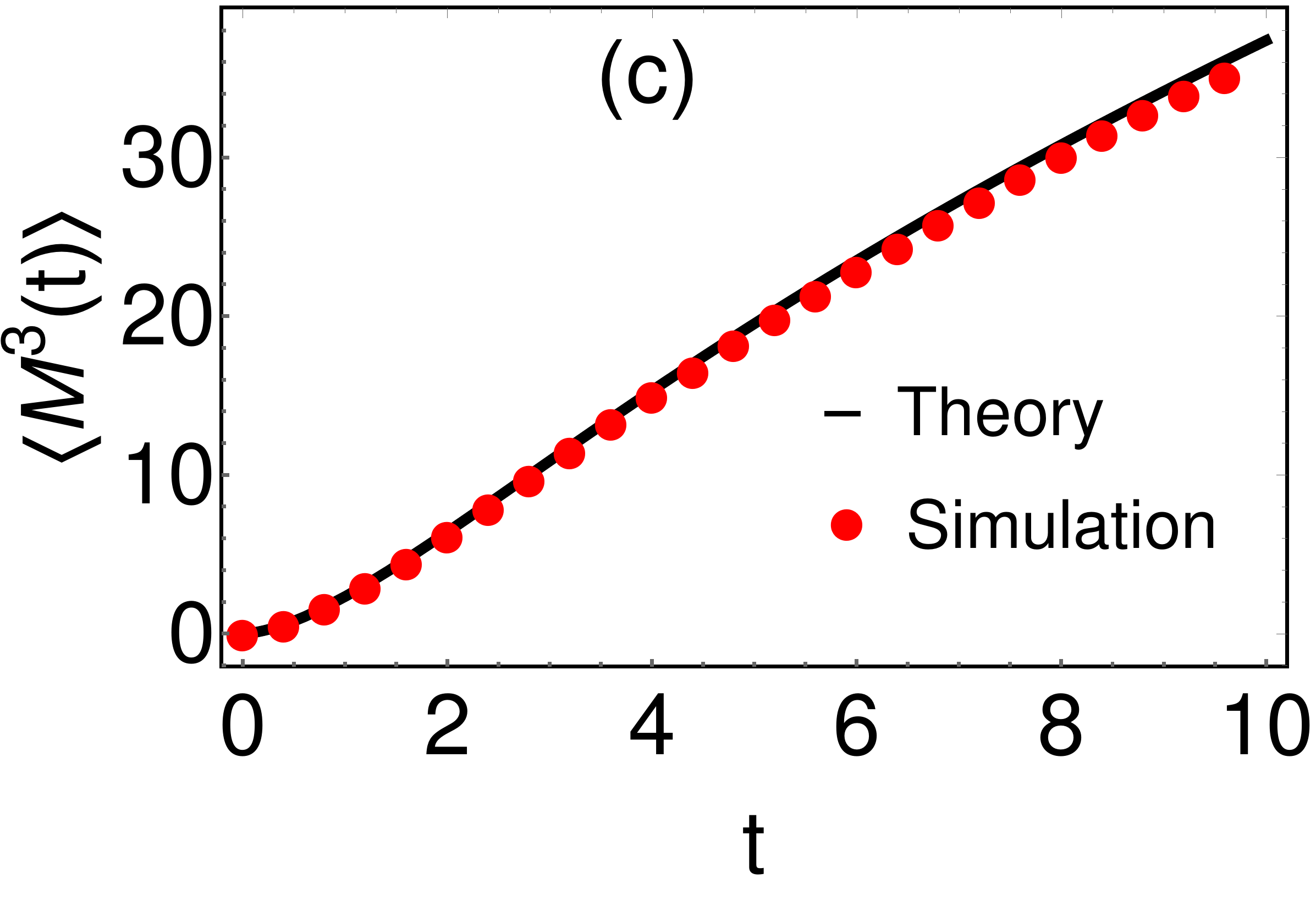}}
\centering
\caption{Comparison of $\langle M^{n}(t) \rangle$ in Eq. \eqref{mom-M-dd} for the drift-diffusion process against the numerical simulations for (a) $n=1,$ (b) $n=2$ and (c) $n=3$. For all the plots, we have chosen $r=1,~v=0.5$.}    
\label{Drifted-BM-max-mom}
\end{figure*} 
\subsection{Diffusion with drift}
\label{DD}

We now consider a particle diffusing in one dimension in presence of a constant drift $v~(\geq 0)$. Herein, our aim is to analyze the statistics of $M$ for this process with dynamics in Eq. \eqref{DDiff}. To this aim, we begin with $\mathcal{Z}_r\left(M, k|s \right)$ in Eq. \eqref{main-eq-5} for which we need the following two quantities \cite{tmax-3}:
\begin{align}
&\bar{S}_0(0, s|M) = \frac{1}{s} \left( 1-e^{- M\left( \sqrt{v^2+2s}-v \right)}\right), \label{BMDeq-1}\\
&\mathcal{Z}_0\left(M, k|s \right) = \frac{\sqrt{v^2+2s}-v}{s} e^{-M\left( \sqrt{v^2+2(s+k)}-v \right)} \label{BMDeq-2}.
\end{align}
Substituting these two equations in Eq. \eqref{main-eq-5} gives us $\mathcal{Z}_r\left(M, k|s \right)$ which can then be suitably used to compute the distribution of $M$ and $t_m$.

Let us first look at the statistics of $M$ for drifted diffusion for which we put $k=0$ in $\mathcal{Z}_r\left(M, k|s \right)$ in Eq. \eqref{main-eq-5} along with $\bar{S}_0(0, s|M) $ and $\mathcal{Z}_0\left(M, k|s \right)$ in Eqs. \eqref{BMDeq-1} and \eqref{BMDeq-2} respectively. This yields  
\begin{align}
\mathcal{Z}_r\left(M, k=0|s \right) &= \frac{ (s+r) [\sqrt{v^2+2(s+r)}-v]}{[s+re^{- M\left( \sqrt{v^2+2(s+r)}-v \right)}]^2} \nonumber \\
& ~~~~~~\times e^{- M\left( \sqrt{v^2+2(s+r)}-v \right)}.
\label{dist-max-BMD-eq-1}
\end{align}
As done for the Brownian motion, we first analyse the moments of $M(t)$ followed by the distribution. Using Eq. \eqref{dist-max-BMD-eq-1}, it is easy to see that the $n$-th order moment is given by
\begin{align}
\int _0 ^{\infty} dt e^{-st} \langle M^{n}(t) \rangle& = \int _{0}^{\infty} dM \mathcal{Z}_r\left(M, k=0|s \right), \label{BMDeq-3} \\
& = -\frac{ n! (s+r)~\text{Li}_n \left(-\frac{r}{s} \right)}{s r[\sqrt{v^2+2(s+r)}-v]^n}~, \label{BMDeq-4}\\
&= -\frac{ n! (s+r)~\sum _{k=1}^{\infty} \frac{(-r)^k}{k^n ~s^{k}}}{s r[\sqrt{v^2+2(s+r)}-v]^n},
\label{BMDeq-5}
\end{align}
where $\text{Li}_n \left(-\frac{r}{s} \right)$ is the PolyLog function and its series representation $\text{Li}_n \left(-y \right) = \sum _{k=1}^{\infty} \frac{(-y)^k}{k^n}$  has been used in Eq. \eqref{BMDeq-4}. The Laplace transform in Eq. \eqref{BMDeq-5} can be inverted exactly as illustrated in the appendix \ref{M-mom-ddiff}. The $n$-th moment obeys the scaling form
\begin{align}
\langle M^n(t)\rangle = \left( \frac{t}{2}\right)^{n/2} \mathbb{H}_n \left( r t, -\frac{v \sqrt{t}}{\sqrt{2}}\right),
\label{mom-M-dd}
\end{align}
where the scaling function $\mathbb{H}_n(z,y)$ is given by
%\begin{widetext}
\begin{align}
&\mathbb{H}_n(z,y)=\frac{(-1)^n n}{z} \sum _{k=1}^{\infty} \frac{(-z)^k}{k^n} \int _{0}^{1} dw e^{-\left( z+y^2\right)w}\frac{(1-w)^{k-1}}{\Gamma(k)} \nonumber \\
&\left[ 1+z \frac{1-w}{k}\right] \left[ \frac{\delta _{n,1}}{\sqrt{\pi w}} - \frac{d^{n-1}}{d y^{n-1}}\left( y e^{y^2 w} \text{Erfc}(y \sqrt{w})\right)\right],
\label{sca-mom-M-dd}
\end{align}
%\end{widetext}
where $\text{Erfc}(z)=1-\text{Erf}(z)$ is the complementary error function.
The scaling function $\mathbb{H}_n(z,y)$ can be simplified further for some values of $n$. For example, one can perform the summation over $k$ for $n=1$ to get
\begin{align}
\mathbb{H}_1(z,y) &= \frac{|y|}{z} \left[ \gamma _E +\Gamma \left(0, z \right) + \log z \right] \nonumber \\
&+ \int _{0}^{1} \frac{dw}{z} \frac{\mathcal{J}_w(z,y)}{(1-w)} \left[1-e^{(1-w)z} \right],~~~\text{with }\\
\mathcal{J}_w(z,y)&=\frac{e^{-(z+y^2)w}}{\sqrt{\pi w}}+\sqrt{z+y^2}~ \text{Erf}\left( \sqrt{(z+y^2)w}\right).
\end{align}
However, for arbitrary $n$, the scaling function is given by Eq. \eqref{sca-mom-M-dd}. In Figure \ref{Drifted-BM-max-mom}, we have plotted the first three moments of $M(t)$ and also compared with the numerical simulations to find an excellent match. Next, we look at the asymptotic forms of $\langle M^n(t)\rangle$ to study the effect of resetting. For small $z$, we consider Eq. \eqref{sca-mom-M-dd} and perform a direct expansion in $z$ to obtain the behaviour of $\mathbb{H}_n(z,y)$. On the other hand, for large $z$, the scaling function goes as $\mathbb{H}_n(z,y) \simeq \log^n z$ and thus, the asymptotic forms read
\begin{align}
\mathbb{H}_n(z,y)  & \simeq n(-1)^{n} \mathbb{C}_n(y) + O(z),~~~~~\text{as }z \to 0, \label{dd-mom-as-eq-1}\\
& \simeq \frac{\log ^n z}{\left(\sqrt{y^2+z}-y \right)^n}~~~~~~~~~~~~\text{as }z \to \infty. 
\label{dd-mom-as-eq-2}
\end{align}
The function $\mathbb{C}_n(y)$ in Eq. \eqref{dd-mom-as-eq-1} is given by
\begin{align}
\mathbb{C}_n(y) &= \int _{0}^{1} dw e ^{-y^2 w} \frac{d^{n-1}}{dy^{n-1}} \left[ y e^{y^2 w} \text{Erfc}(y \sqrt{w})\right],\label{dd-mom-as-eq-3}\\
&~~~~~~~~~~~~~~~~~~~~~~~~~~~~~~~~~~~~~~~~~~~\text{for } n \neq 1, \nonumber\\
&=y ~\text{Erfc}(y)-\frac{e^{-y^2}}{\sqrt{\pi}} - \frac{\text{Erf}(y)}{2 y} ,~~~\text{for }n=1.
\label{dd-mom-as-eq-4}
\end{align}
Inserting the forms of $\mathbb{H}_n(z,y)$ from Eqs. \eqref{dd-mom-as-eq-1} and \eqref{dd-mom-as-eq-2} in Eq. \eqref{mom-M-dd}, we find that $\langle M^n(t)\rangle$ has the following asymptotic forms
\begin{align}
\langle M^n(t)\rangle &= (-1)^n n \mathbb{C}_n \left(-v\sqrt{\frac{t}{2}} \right) t^{n/2} +O(r t), \nonumber \\
&~~~~~~~~~~~~~~~~~~~~~~~~~~~~~~~~~\text{for }t \ll r^{-1}, \label{dd-mom-as-eq-5}\\
& = \frac{\log ^n rt}{\left(\sqrt{v^2+2 r}-v \right)^n} + O\left(\log ^{n-1}r t \right), \nonumber \\
&~~~~~~~~~~~~~~~~~~~~~~~~~~~~~~~~~\text{for }t \gg r^{-1}. \label{dd-mom-as-eq-6}
\end{align}
At leading order, $\langle M^n(t)\rangle$ for $t \ll r^{-1}$ matches with the drift-diffusion without resetting \cite{tmax-3}. Also, for $t \gg r^{-1}$ we find that the maximum scales logarithmically with time as $M(t) \sim \log (rt)$. Thus even though the position density reaches a non-equilibrium steady state at late times \cite{Restart4}, the maximum $M(t)$ keeps growing albeit slowly in logarithmical scale. This behavior can also be  understood heuristically from the extreme value statistics. To illustrate this, we look at the distribution of $M(t)$ at $t \gg r^{-1}$ for which we analyse $\mathcal{Z}_r\left(M, k=0|s \right)$ in Eq. \eqref{dist-max-BMD-eq-1} for $r \gg s$. Approximating $s+r \simeq r$ in Eq. \eqref{dist-max-BMD-eq-1}, we get
\begin{align}
\mathcal{Z}_r\left(M, k=0|s \right) \simeq & \frac{ r [\sqrt{v^2+2r}-v]e^{- M\left( \sqrt{v^2+2r}-v \right)}}{[s+re^{- M\left( \sqrt{v^2+2r}-v \right)}]^2}.
\label{dd-mom-as-eq-7}
\end{align}
Performing the inverse Laplace transformation with respect to $s$ gives the distribution of $M$ at large $t$ as 
\begin{align}
P_r(M|t)= rt \alpha e^{-M \alpha} ~\text{exp}\left(- r t e^{-M \alpha} \right),~~~~\text{for }t\gg r^{-1},
\label{dd-mom-as-eq-8}
\end{align}
which is a Gumbel distribution and $\alpha = \sqrt{v^2+2r}-v$. In \fref{driftBM-dist} (top panel), we have compared the distribution $P_r(M|t)$ in Eq. \eqref{dd-mom-as-eq-8} with the same obtained from the numerical simulations. We observe an excellent agreement. Note that using this form of $P_r(M|t)$, it is straightforward to reproduce the $n$-th moment as given in Eq. \eqref{dd-mom-as-eq-6}. Again, the appearance of the Gumbel distribution for $M(t)$ in \eref{dd-mom-as-eq-8} can be understood from the EVS as done for the simple diffusion.

\section{Statistics of arg-max $t_m$}
\label{statistics-of-argmax}
In this section, we will present the results for statistics of the arg-max i.e., the time $t_m$ at which the maximum $M$ occurs. The starting point would be again to consider the joint distribution $\mathcal{Z}_r(M,k|s)$ of $M$ and $t_m$ in Laplace space given in Eq. \eqref{main-eq-5}. Next, we would integrate out $M$ to obtain an expression for the marginal distribution $\mathcal{Z}_r(k|s)$. We will analyze this quantity to characterize $t_m$ for simple diffusion and then for drift-diffusion process.\\

\subsection{Simple diffusion}
\begin{figure*}[t]
  \centering
  \subfigure{\includegraphics[scale=0.28]{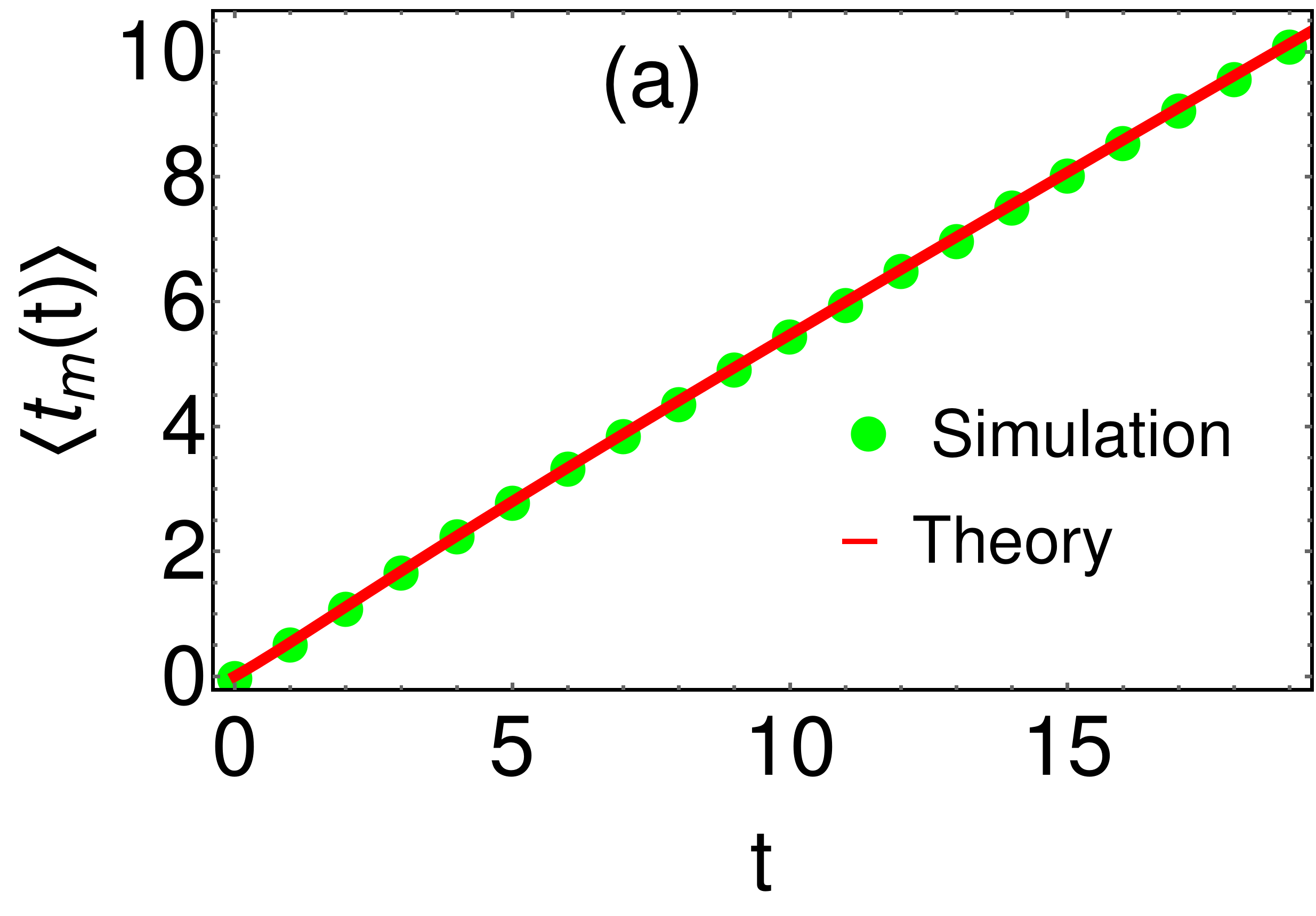}}
  \subfigure{\includegraphics[scale=0.284]{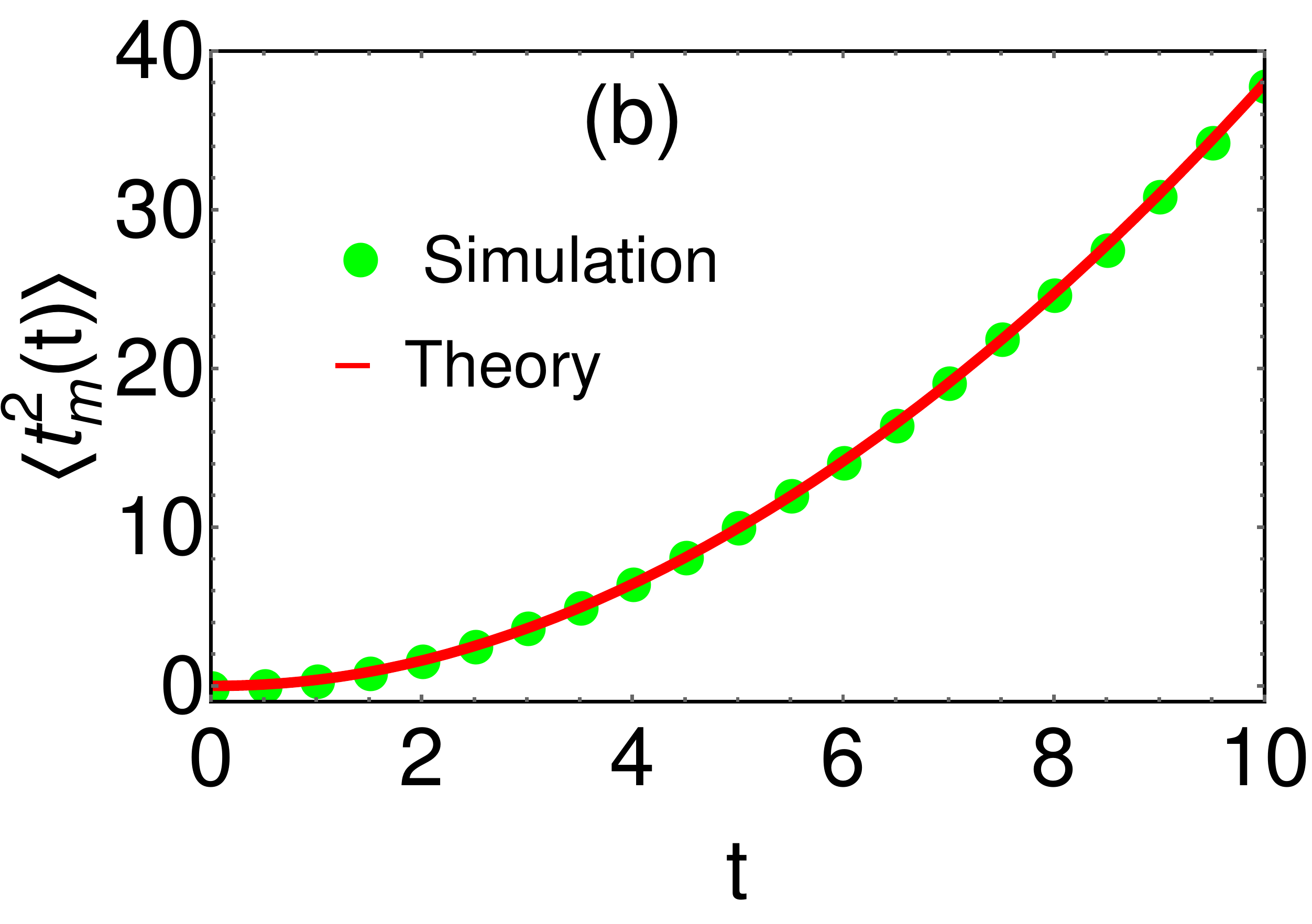}}
  %\subfigure{\includegraphics[scale=0.22]{drift-max-mom-3-BM.pdf}}
\centering
\caption{Comparison of the first two moments of $t_{m}$ in Eqs. \eqref{BMeq-18} and \eqref{BMeq-20} against the same obtained from the numerical simulations for the simple diffusion. We have set $r=1$ for the simulation purpose.}
  \label{BM-tmax-mom}
\end{figure*}

We now look at the statistical properties of $t_m$ for Brownian motion with resetting. For free Brownian motion, the distribution of $t_m$ is $P_0(t_m|t) = \frac{1}{\pi \sqrt{t_m(t-t_m)}}$ whose cumulative exhibits the `Arc-sine' law (see \eref{arc-sine}). To investigate how the statistical properties are influenced due to resetting, we first consider $\mathcal{Z}_r\left(M, k|s \right) $ in Eq. \eqref{main-eq-5} and integrate it over $M$. The resultant function $\mathcal{Z}_r(k|s) =  \int _0 ^{\infty} d M \mathcal{Z}_r\left(M, k|s \right)$ gives the double Laplace transformation of the distribution $P_r(t_m|t)$ with respect to $t_m~(\to k)$ and $t~(\to s)$. Inserting $\mathcal{Z}_0\left(M, k|r+s \right)$ and $\bar{S}_0(0, s|M)$ from Eqs. \eqref{BMeq-1} and \eqref{BMeq-2} in $\mathcal{Z}_r\left(M, k|s \right) $ in Eq. \eqref{main-eq-5} and performing the integration over $M$, we get
\begin{align}
\mathcal{Z}_{r} \left(k|s \right) =& \int _{0}^{\infty} dM\left[\frac{ \sqrt{2(r+s)} e^{-\sqrt{2(r+s+k)}M}}{\left(s+re^{-\sqrt{2(r+s)}M}\right)} \right.\nonumber\\
&\left.  \times \frac{(r+s+k)}{\left( s+k+r e^{-\sqrt{2(r+s+k)}M} \right)}\right].
\label{BMeq-12}
\end{align}
Doing a change of variable: $w = e^{-\sqrt{2(r+s+k)}M}$ in the above equation gives
\begin{align}
\mathcal{Z}_{r} \left(k|s \right) = \int _{0}^{1}dw \frac{\sqrt{(r+s)(r+s+k)} }{\left(s+r w^{\phi} \right) \left( s+k + r w\right)},
\label{BMeq-13}
\end{align}
where $\phi = \sqrt{\frac{r+s}{r+s+k}}$. We next use $\mathcal{Z}_{r} \left(k|s \right)$ from Eq. \eqref{BMeq-13} to compute the moments of $t_m$ explicitly at all time. The $n$-th order moment $\langle t_m^n(t) \rangle$ can be written in terms of $\mathcal{Z}_{r} \left(k|s \right)$ as
\begin{align}
\int _{0}^{\infty} dt ~e^{-s t}~\langle t_m^n(t) \rangle = (-1)^n\left[\frac{d^n}{d k^n}\mathcal{Z}_{r} \left(k|s \right) \right] _{k=0}.
\label{mom-tm-eq-46}
\end{align}
Inserting $\mathcal{Z}_{r} \left(k|s \right)$ from Eq. \eqref{BMeq-13} in Eq. \eqref{mom-tm-eq-46}, one can analyse the individual moments of $t_m(t)$ although getting a closed form for the $n$-th order moment turns out to be difficult. Below, we provide the exact expression for the first two moments of $t_m$:
\begin{widetext}
\begin{align}
\langle t_{m}(t) \rangle &= \frac{1}{4 r} \left[ 2 r t+e^{-r t}-1+ \gamma _E +\Gamma \left( 0, r t\right) + \log (rt)\right]
\label{BMeq-18}, \\
\langle t_{m}^2(t) \rangle &= \frac{1}{24 r^2} \Bigg[ e^{-r t}\left(9+5 r t \right)-9+4 rt \left(1+2 rt \right) +4-3 \gamma _E-3 \Gamma \left( 0, r t\right)-3 \log \left( rt\right)\Bigg. \nonumber \\
& \left. + 4 r t \Big\{-1+\gamma _E +\Gamma \left(0, rt \right)+\log(r t) \Big\} +e^{-r t} \Big\{-4-3 \gamma _E + 3 \text{Chi}(rt)+3 \text{Shi}(rt)-3 \log (rt) \Big\}\right. \nonumber \\
& \Bigg. +2 e^{- r t} \int _0^{ r t}~ dy~ y ~e^{y}~_3F_3 \Big(\{1,1,1 \},\{2,2,2 \},\{-r t \} \Big)\Bigg],
\label{BMeq-20}
\end{align}
\end{widetext}
where $\text{Chi}(z)$ and $\text{Shi}(z)$ stands for hyperbolic cosine integral and hyperbolic sine integral respectively and $~_3F_3 \Big(\{1,1,1 \},\{2,2,2 \},\{z\} \Big)$ is the generalised hypergeometric function \cite{Table}. Expressions of  $\langle t_{m}(t) \rangle$ and $\langle t^2_{m}(t) \rangle$ are derived in \aref{appen-tm-BM}. In Figure \ref{BM-tmax-mom}, we have plotted the first two moments of $t_m(t)$ and compared them against the same obtained from the numerical simulations. An excellent match is observed. 

\begin{figure}[b]
\centering
  \includegraphics[scale=0.25]{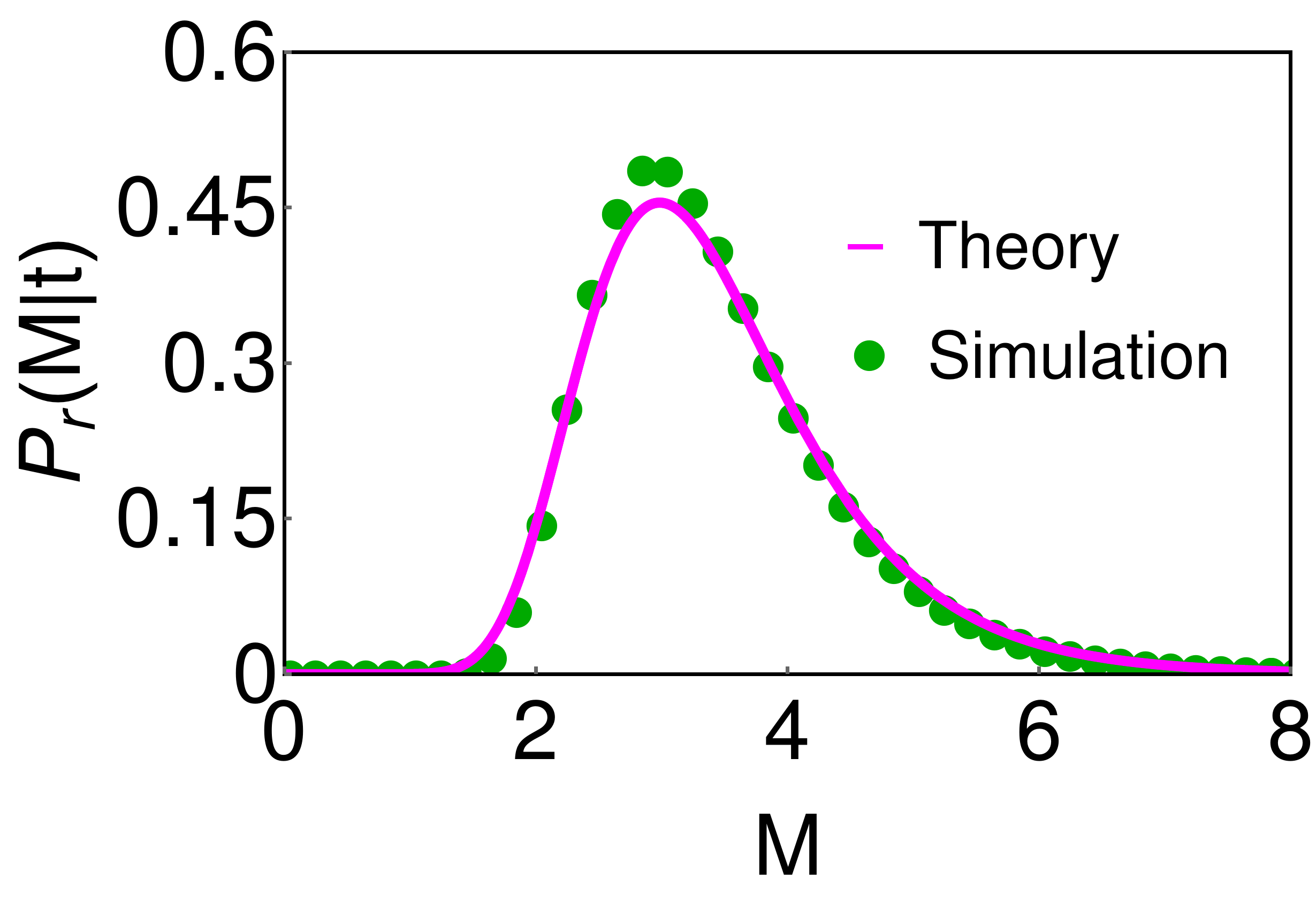}
  \includegraphics[scale=0.25]{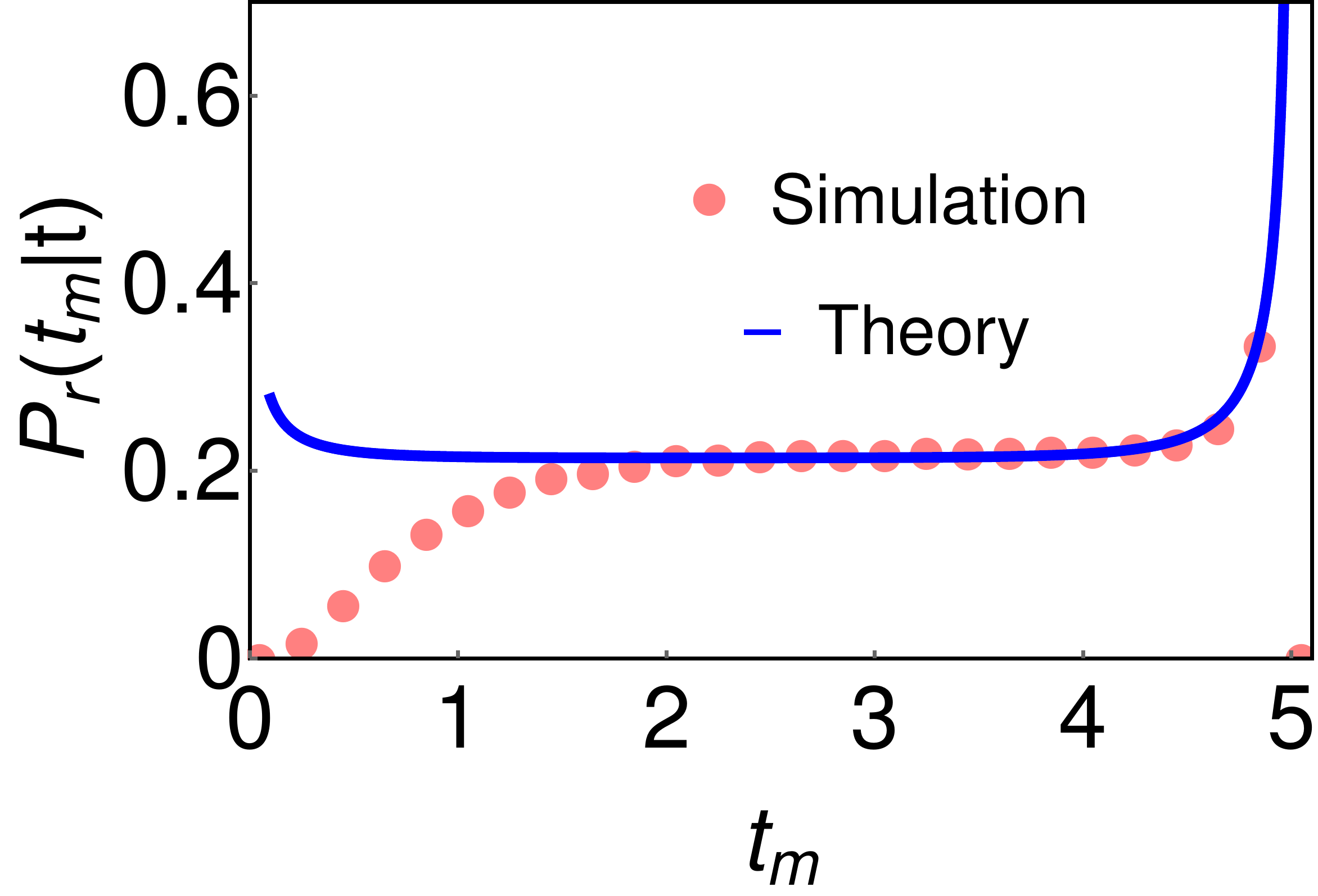}
  \caption{\textit{Top panel:} comparison of the distribution $P_r(M|t)$ in Eq. \eqref{dd-mom-as-eq-8} for the drift-diffusion process against the numerical simulations. For this plot, we have taken: $r=2,~v=1$ and $t=20$. \textit{Bottom panel:} here we have compared the distribution $P_r(t_m|t)$ in Eq. \eqref{density-arg-max-DD} for the drift-diffusion to the numerical simulations. As before, the data fits the theory perfectly in the large $t_m$ limit. Parameters chosen for this simulation: $r=1.5$ and $t=5$.}
  \label{driftBM-dist}
\end{figure}

To analyse the consequences of resetting, we study the moments of $t_m(t)$ at short and large times. It turns out that for these times, one can write a closed form for the $n$-th order moment. Starting from $\mathcal{Z}_{r} \left(k|s \right)$ in Eq. \eqref{BMeq-13}, we perform simplifications for the cases $s r^{-1} \to 0$ and $s r^{-1} \to \infty$ which in the time domain correspond to $t \gg r^{-1}$ and $t \ll r^{-1}$ respectively. We then use Eq. \eqref{mom-tm-eq-46} to compute the moments for these cases. In order to avoid discontinuity of the presentation, we have relegated these detailed calculations to appendix \ref{appen-asy-tm-BM} and present only the final results here. The moments read 
\begin{align}
\langle t_{m}^n (t) \rangle 
& \simeq \frac{(2n-1)! t^{n}}{n! (n-1)! 2^{2n-1}}  + \mathcal{A}_n\gamma t^{n+1},  ~\text{for } t \ll \frac{1}{r}
\label{BMeq-16-new}\\
&\simeq \frac{t^n}{n+1} + \frac{t^{n-1} \log (rt)}{2r(n+1)}, ~~~~~~~~~~\text{for } t \gg \frac{1}{r}, \label{BMeq-16}
\end{align}
where $\mathcal{A}_n$ in the first line is given in Eq. \eqref{An-exp}. For $t \ll \frac{1}{r}$, we recover the results of the free Brownian motion \cite{Levy}. Moreover, we observe that the leading order scaling of the $n$-th order moment with respect to $t$ for both large and small times are same i.e. $\langle t_{m}^n (t) \rangle \sim t^{n}$. However, the corresponding prefactors are different. As we illustrate later that the leading order behaviour of $\langle t_{m}^n (t) \rangle$ at $t \gg \frac{1}{r}$ is independent of the underlying stochastic process as long as every reset event renews the process. Contrarily, the sub-leading terms are sensitive to the underlying stochastic process and resetting rate.

In the remaining part of this section, we analyse the distribution of $t_m$ which we denote by $P_r(t_m|t)$. To compute this distribution, we have to perform the double inverse Laplace transformation of $\mathcal{Z}_{r} \left(k|s \right)$ in Eq. \eqref{BMeq-13} with respect to $k$ and $s $. Performing inversion for arbitrary values of $s$ and $k$ turns out to be difficult. However, one can make some analytic progress by analysing $\mathcal{Z}_{r} \left(k|s \right)$ in various limits of $k r^{-1}$ and $s r^{-1}$. For $r \gg s$ and $r \gg k$, we approximate $\phi \simeq 1$ in the expression of $\mathcal{Z}_{r} \left(k|s \right)$ in Eq. \eqref{BMeq-13} to get
\begin{align}
\mathcal{Z}_{r} \left(k|s \right)\simeq \int _{0}^{1}dw \frac{\sqrt{(r+s)(r+s+k)} }{\left(s+r w \right) \left( s+k + r w\right)}.
\label{dist_new_eqq_573}
\end{align}
Fortunately, one can now invert this double Laplace transform to get the distribution $P_r(t_m|t)$ [see appendix \ref{appen-dist-BM}]. Performing the inverse Laplace transformation, we get 
\begin{align}
P_r(t_{m}|t) &\simeq e^{-r t } \frac{\left[g(r t_m)+g(r(t-t_m))-1 \right]}{\pi \sqrt{t_m(t-t_m)}} \nonumber\\
&+ r e^{-r t} \int _0^{1} ~dw ~w ~e^{r t w}  ~ \text{Erf} \left(\sqrt{r w t_{m}} \right) \nonumber \\
&~~~~~~~~~\times  \text{Erf} \left(\sqrt{r w (t-t_{m})} \right),
\label{BMeq-15-new}
\end{align} 
where $g(z) = ~_2F_2 \left( \{ 1,1 \}, \{1/2,2 \},z \right)$. Note that this expression is valid only in the limits $rt \to \infty$ and $r t_m \to \infty$. In Fig. \ref{BM-dist} (bottom panel), we have plotted $P_r(t_{m}|t) $ and compared it against the numerical simulations. While our analytic result is consistent with the simulation data at large $t_m$, the match is poor at small $t_m$. This deviation stems from the fact that $\mathcal{Z}_{r} \left(k|s \right)$ in Eq. \eqref{dist_new_eqq_573} is valid only for $r \gg k$ which translates to $t_m \gg r^{-1} $ in the time domain.

We now look at $P_r(t_{m}|t)$ when $t_m \neq t$ for which \eref{BMeq-15-new} can be simplified further. Approximating $_2F_2 \left( \{ 1,2 \}, \{1/2,2 \},z \right) \simeq \sqrt{\frac{\pi}{z}} e^{z}$ and $\text{Erf}(z) \simeq 1$ for $z \to \infty$ and using them in \eqref{BMeq-15-new}, we find that the leading order behaviour of $P_r(t_{m}|t) \simeq \frac{1}{t}$. To get the sub-leading terms of $P_r(t_{m}|t) $, one has to consider higher order terms on $r s^{-1}$ while performing simplifications in $\mathcal{Z}_{r} \left(k|s \right)$ in Eq. \eqref{dist_new_eqq_573}. We refer to appendix \ref{appen-asy-tm-BM} [see Eq. \eqref{appen-asy-tm-BM-eq-5}] for the derivation of the sub-leading corrections and present the results here such that 
%The simplified form of $P_r(t_{m}|t) $ for $rt \to \infty$ and $r t_m \to \infty$ and $ r(t-t_m) \to \infty $ reads
\begin{align}
P_r(t_{m}|t) \simeq \frac{1}{t} + \frac{\log(r t)}{ 2r t^2}.
\label{BMeq-15}
\end{align} 
Using this form of $P_r(t_{m}|t)$, it is easy to verify that the $n$-th order moment of $t_m(t)$ is indeed given by Eq. \eqref{BMeq-16}. It is worth remarking that the $\frac{1}{t}$ form of $P_r(t_{m}|t) $ is quite different than the form of distribution of $t_m$ without reset which is given by  $P_0(t_m|t) = \frac{1}{\pi \sqrt{t_m(t-t_m)}}$. We later show that $\frac{1}{t}$ form of $P_r(t_{m}|t) $ under resetting is independent of the underlying stochastic process as long as the process forgets about its prior history after every reset event.

\subsection{Diffusion with drift}
\label{dBM}
We start again by inserting $\mathcal{Z}_0(M,k|r+s)$ and $\bar{S}_0(0, s|M)$ for drift diffusion process from Eqs. (\ref{BMDeq-1})-(\ref{BMDeq-2}) in \eref{main-eq-5} to obtain the following expression for $\mathcal{Z}_r(M,k|s)$ given by
\begin{align}
\mathcal{Z}_{r} \left(M,k|s \right) = \frac{\mathcal{B}_{s+r}(r+s+k) ~e^{- M \mathcal{B}_{s+r+k}}}{ \left[s+re^{- M \mathcal{B}_{s+r}}\right] \left[s+k+re^{- M\mathcal{B}_{s+k+r}}\right]},
\label{BMDeq-12-v}
\end{align}
where $\mathcal{B}_{s} = \sqrt{v^2+2s}-v $. Now performing the integration over $M$ and after some algebraic simplifications, we find
\begin{align}
\mathcal{Z}_{r} \left(k|s \right) &=\frac{\sqrt{v^2+2(s+r)}-v}{\sqrt{v^2+2(s+k+r)}-v} \nn \\
&\times \int _{0}^1~dw~\frac{(r+s+k)}{(s+r w^{\phi _d })(s+k+rw)}~,
\label{BMDeq-13-v}
\end{align}
with $\phi _d=\frac{\sqrt{v^2+2(s+r)}-v}{\sqrt{v^2+2(s+k+r)}-v}$. We can now use \eref{mom-tm-eq-46} to get the expression for the moments in the Laplace space. However, the inversion process to obtain the moments in real time becomes quite tedious. Here, we just present the first moment for brevity. The first moment in Laplace space reads (using \eref{BMDeq-13-v} in \eref{mom-tm-eq-46})

\begin{figure}[t]
\centering
  \includegraphics[scale=0.25]{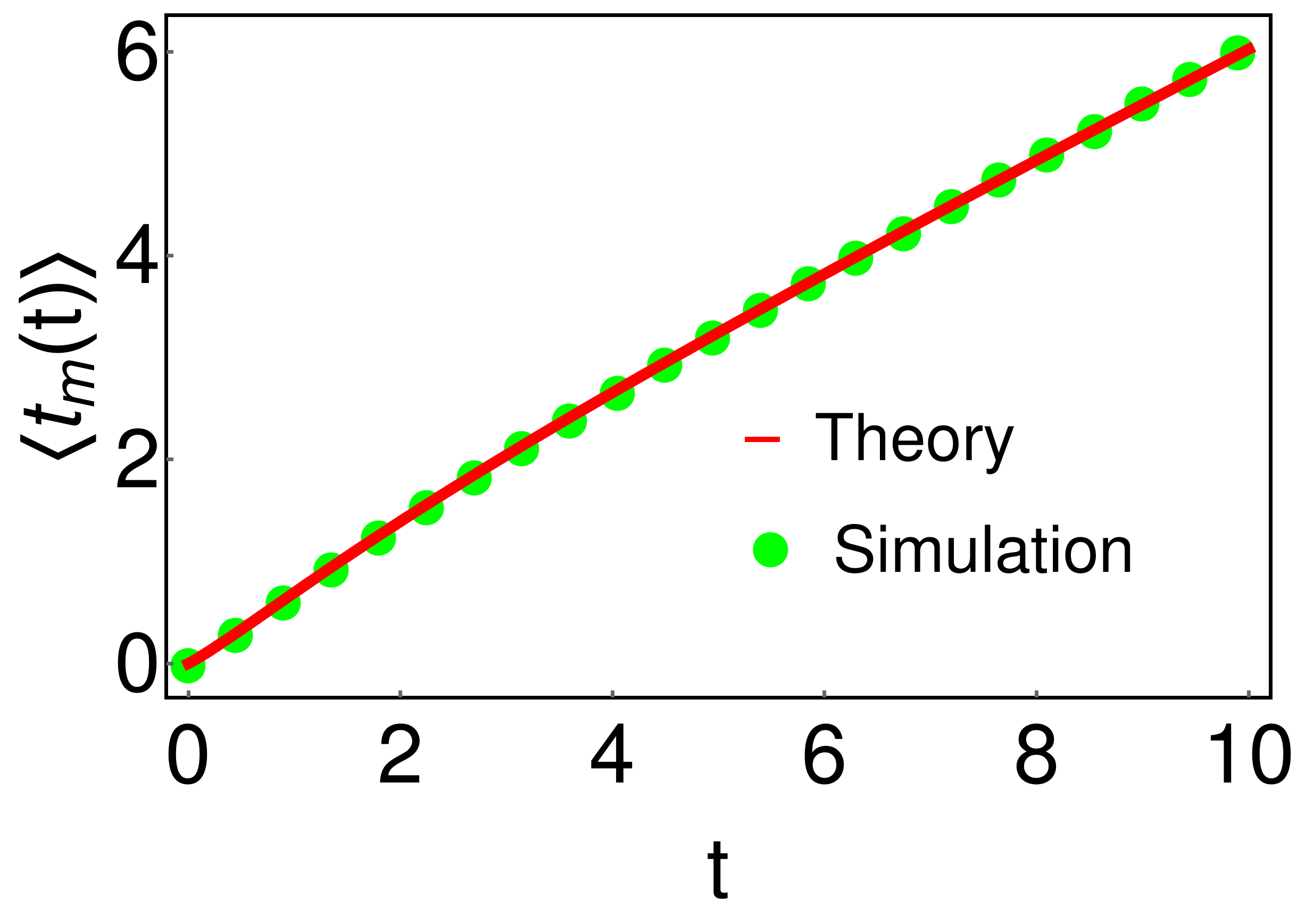}
  \caption{Comparison of $\langle t_m(t) \rangle$ in Eq. \eqref{mean-tm-dd-new} with the numerical simulations. We have chosen $r=1$ and $v=1$ for the simulation purpose.}
  \label{dd-mean-tm-pic}
\end{figure}

\begin{align}
      &\int_0^\infty~dt~ e^{-st} \langle t_m(t) \rangle = \tilde{I}_1(s)+\tilde{I}_2(s),~~~~\text{with} \label{mean-dd-tm}\\
      &\tilde{I}_1(s) =  \frac{2r+s+s \frac{v}{\sqrt{v^2+2(r+s)}}}{4s^2(r+s)}, \\
      & \tilde{I}_2(s) = \left[ 1+\frac{v}{\sqrt{v^2+2(r+s)}}  \right] \frac{1}{4rs} \log \left( \frac{r+s}{s} \right).
\end{align}
To get $\langle t_m(t) \rangle$, we use the following inverse Laplace transformation:
\begin{widetext}
\begin{align}
    I_1(t)&=\mathcal{L}_{s \to t}^{-1} \left[ \tilde{I}_1(s) \right]=\frac{1}{4r} \left( 2rt+e^{-rt} \text{Erfc}\left[  \frac{v\sqrt{t}}{\sqrt{2}}\right]+\frac{v}{\sqrt{v^2+2r}} \text{Erf}\left[ \frac{\sqrt{(v^2+2r)t}}{\sqrt{2}} \right]-1 \right),\\
    I_2(t)&=\mathcal{L}_{s \to t}^{-1} \left[ \tilde{I}_1(s) \right]= J(t)+\int_0^t~d\tau~J(\tau)~\frac{v e^{-\frac{v^2+2r}{2}(t-\tau)}}{\sqrt{2 \pi (t-\tau)}}, \\
  \text{where~~}  J(t)&=\frac{1}{4 r} \left[  \gamma _E +\Gamma \left( 0, r t\right) + \log (rt)\right].
\end{align}
\end{widetext}
Inserting these inverse Laplace transforms in Eq. \eqref{mean-dd-tm}, we get
\bea
\langle t_m(t) \rangle =I_1(t)+I_2(t).
\label{mean-tm-dd-new}
\eea
One can also proceed to compute the higher moments in a similar manner, but the expressions are quite involved. Hence, we do not present them here. In Figure \ref{dd-mean-tm-pic}, we have compared $\langle t_m(t) \rangle$ in Eq. \eqref{mean-tm-dd-new} with the numerical simulations to find a perfect agreement between them.

We now turn our attention to analyze $P_r(t_m|t)$ for the drift-diffusion process. Similar to the simple diffusion, performing an inversion for arbitrary values of $s$ and $k$ turns out to be difficult. Thus, we make the approximations $r\gg s$ and $r \gg k$ to have $\phi _d \simeq 1$, and \eref{BMDeq-13-v} simplifies to
\begin{align}
\mathcal{Z}_{r} \left(k|s \right)  &\simeq \frac{\sqrt{v^2+2(s+r)}-v}{\sqrt{v^2+2(s+k+r)}-v}~ \nn \\
&\times \int _{0}^{1}~dw~ \frac{(r+s+k)}{\left(s+r w \right) \left( s+k+r w\right)}.
\label{BMDeq-14-v}
\end{align}
It is possible to perform the double Laplace inversion (see \aref{Appendix-drift-diffusion-tmax-PDF}) to eventually arrive at
\bea
P_r(t_m|t)\simeq \mathbb{I}_1(t_m,t)+\mathbb{I}_2(t_m,t)+\mathbb{I}_3(t_m,t),
\label{density-arg-max-DD}
\eea
where $\mathbb{I}$-functions are given in Eqs. (\ref{def-I1}-\ref{def-I3}). Note again that this expression is only valid in the limits of $rt \to \infty$ and $rt_m \to \infty$. We verify this result in Fig \ref{driftBM-dist} (bottom panel). One can again approximate the above expression in large time (like we have shown in the case of simple diffusion) to find the leading order behavior: $P_r(t_m|t) \simeq \frac{1}{t}$, which is again independent of the process details. However, one would expect that there will be some correction (sub-leading) terms to this leading behavior. To find them, we first perform the small $k$ expansion of $\phi _d$ to obtain
\bea
\phi _d &\simeq &1-b(r,v)k~, \nn \\
~\text{with}~ b(r,v)&=&\frac{1}{\sqrt{v^2+2r}(\sqrt{v^2+2r}-v)}.
\label{b(r,v)}
\eea
Further, we approximate $w^{\phi _d } =e^{\phi _d \log w}\simeq e^{(1-bk)\log w} \simeq w(1-bk \log w)$.
Moreover, we also take the large $t$, large $t_m$ limit so that $s+r \simeq r$ and $s+r+k \simeq r$
and thus
\begin{align}
\mathcal{Z}_{r} \left(k|s \right) =r~\int _{0}^1~\frac{dw}{(s+r w-rbwk \log w)(s+k+rw)}.
\end{align}
We now first take the inverse Laplace transform with respect to $k \to t_m$, and then we take the inverse Laplace transform with respect to $s \to t$ to find
\begin{align}
P_r(t_m|t) =\int _{0}^1~dw\frac{re^{-rtw}}{1+brw \log w} \left[ 1-\Theta \left( t+\frac{t_m}{brw \log w} \right) \right].
\label{theta-drift}
\end{align}
Let us now do the following transform $rtw=y$ and then
we have
\begin{align}
w \log w = \frac{y}{rt} \log \frac{y}{rt}\simeq -\frac{y}{rt} \log rt ~~~~\text{for}~~~rt \gg 1~.
\end{align}
The argument inside $\Theta$ function in \eref{theta-drift} becomes negative, and hence this term does not contribute. Eventually, we are left with
\bea
P_r(t_m|t)= r ~\int _{0}^{rt}~\frac{dy}{rt}~\frac{e^{-y}}{1-a(r,v)y}~,
\eea
where $a(r,v)=b(r,v)r~ \frac{\log rt}{rt} \ll 1~~\text{when~~} rt \gg 1$. 
Expanding in small $a$ argument and also setting the upper limit to $+\infty$, we get
\begin{align}
P_r(t_m|t) \simeq &\frac{1}{t}+\frac{a(r,v)}{t}, \nonumber \\
\simeq & \frac{1}{t}+b(r,v)~\frac{\log rt}{t^2},
\label{dist-tm-dd-large}
\end{align}
where $b(r,v)$ is given by \eref{b(r,v)}. Comparing this expression with that of the simple diffusion in Eq. \eqref{BMeq-15}, we see that even though the leading order behaviour of $P_r(t_m|t)$ in both cases is $1/t$, the sub-leading terms are rather different. The subleading term in Eq. \eqref{dist-tm-dd-large} depends on the drift $v$ (limit for simple diffusion can be checked easily by noting $b(r,v)=1/2r$). Finally, the limiting distribution in Eq. \eqref{dist-tm-dd-large} also gives the moments at large $t$ which read
\bea
    \langle t_m^n \rangle \simeq \frac{t^n}{1+n}+\frac{b(r,v) t^{n-1} \log rt}{r(1+n)}~~,~~t \gg \frac{1}{r},
\eea
where again the logarithmic correction to the universal form of the moments is observed. So far, we have presented rigorous results for the statistics of  $t_m$  for two canonical models namely the simple diffusion followed by the drift-diffusion process. The large time universal form $P_r(t_m|t) \simeq \frac{1}{t}$ is noteworthy in both 
cases. In the following section, we present general arguments to show why this is indeed a robust and universal characteristic for the arg-max for generic stochastic processes. 

\section{Limiting distribution of $t_m$ for general stochastic process}
\label{gen-process}
In the previous sections, we observed that the density of arg-max converges to a uniform distribution of the form $P_r(t_m|t) \simeq 1/t$ when $t, t_m \gg r^{-1}$ and $t_m \neq t$ for both diffusion and drifted diffusion. Naturally, the question arises what are the ramifications of resetting to other stochastic processes. Here, we show that the $1/t$-form of $P_r(t_m|t)$ is completely universal and independent of the underlying stochastic process as long as the process starts afresh after every reset event. To prove this, we analyse our main formula for $\mathcal{Z}_r\left(M, k|s \right)$ in Eq. \eqref{main-eq-5} in the limit of large $rt$. Note that the survival probability $S_0(0,M|t)$ for any stochastic process is bounded as $0 \leq S_0(0,M|t) \leq 1$ for all $t$. This bound suggests that the Laplace transform $\bar{S}_0(0,M|s)$ can be written as
\begin{align}
\bar{S}_0(0,M|s) = \frac{1}{s} -\frac{1}{s}~ \mathcal{U}(M,s),
\label{gen-eq-1}
\end{align}  
where $\mathcal{U}(M,s)$ is a general function with the constraints that we discuss in the following. Since $\bar{S}_0(0,M|s) \geq 0$, we have $\mathcal{U}(M,s) \leq 1$. Furthermore $\bar{S}_0(0,M|s) \leq 1/s$, which in terms of $\mathcal{U}(M,s)$ becomes $\mathcal{U}(M,s) \geq 0$. Also, we expect that the particle will survive without getting absorbed at $x=M\to \infty$ which implies $\bar{S}_0(0,M \to \infty|s) \simeq 1/s $ or equivalently $\mathcal{U}\left(M \to \infty,s \right) \simeq  0$. In what follows, we use Eq. \eqref{gen-eq-1} along with these constraints to analyze $\mathcal{Z}_r\left(M, k|s \right)$ given in Eq. \eqref{main-eq-5}.

\begin{figure}[t]
\includegraphics[scale=0.3]{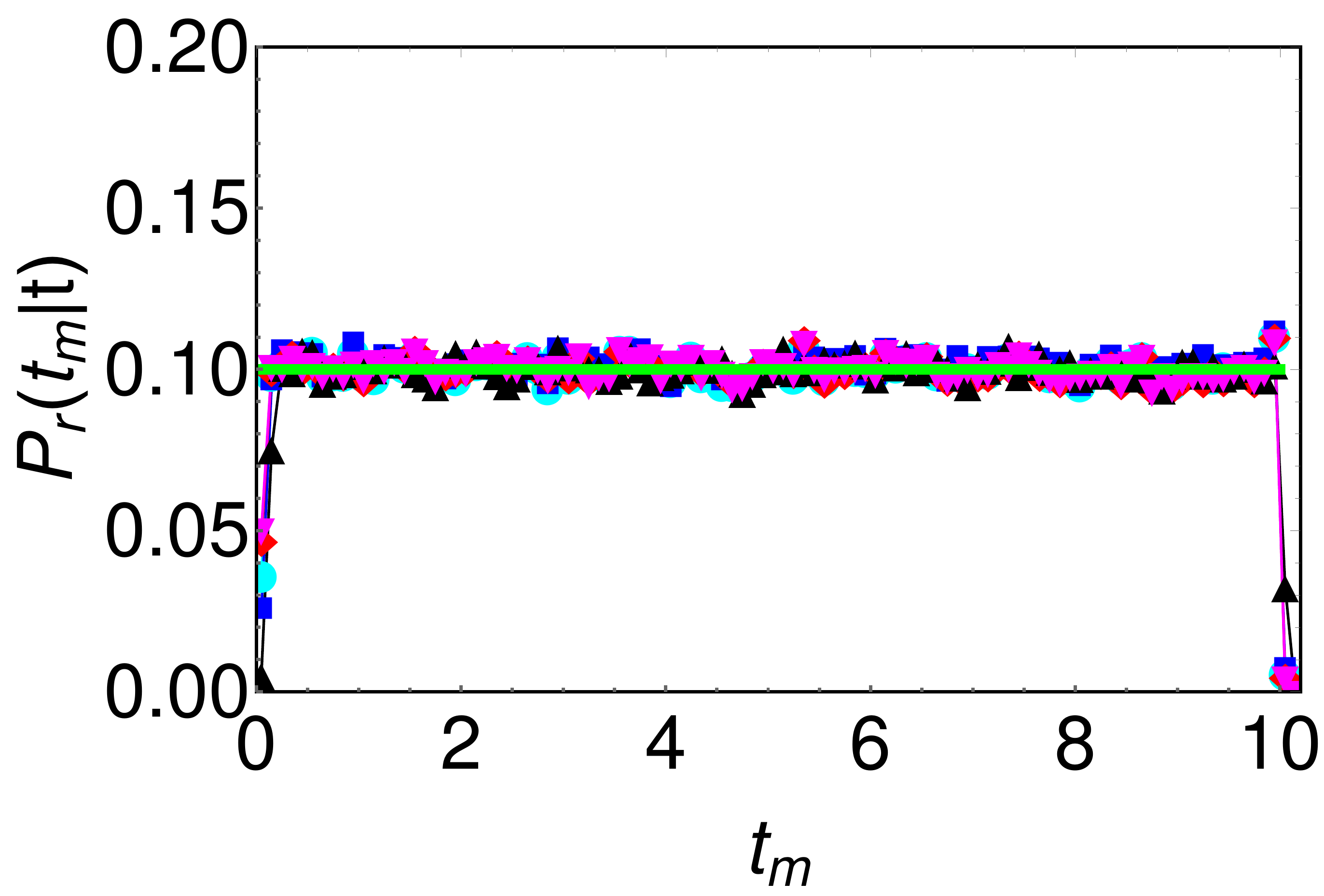}
\centering
\caption{Comparison of $P_r(t_m|t)$ in Eq. \eqref{gen-eq-6} (shown by solid green line) with the simulation results for five different underlying stochastic processes, namely (i) diffusion (cyan), (ii) diffusion with positive drift (blue), (iii) diffusion with negative drift (red), (iv) random acceleration (black) and (v) Ornstein-Uhlenbeck process (magenta). While (i)-(iii) \& (v) are Markov processes, (iv) is a non-Markov one. Simulations are conducted for the parameters $t=1$ and $r=20$ in all the cases (see \sref{simulations} for more details on the model systems).}
\label{dist-tm-all}
\end{figure}

Inserting $\bar{S}(0,M|s) $ from Eq. \eqref{gen-eq-1} in Eq. \eqref{main-eq-5}, we get
\begin{align}
\mathcal{Z}_r\left(M, k|s \right) =\frac{ (r+s)(r+s+k)\mathcal{Z}_0\left(M, k|r+s \right)}{\left[s+r \mathcal{U}(M,r+s)\right] \left[s+k+r \mathcal{U}(M,r+s+k)\right]}.
\label{gen-eq-2}
\end{align}
For $s \to 0$ and $k \to 0$, we approximate $\mathcal{Z}_r\left(M, k|s \right)$ as
\begin{align}
\mathcal{Z}_r\left(M, k|s \right) \simeq \frac{r^2}{\left[s+r \mathcal{U}(M,r)\right] \left[s+k+r \mathcal{U}(M,r)\right]}~\frac{d\bar{S}_0(0,M|r) }{dM},
\label{gen-eq-3}
\end{align}
where we have used $\mathcal{Z}_0 \left(M, k=0|r \right)=\frac{d\bar{S}_0(0,M|r) }{dM}$. This stems from the fact that $S(0,M|t)$ is the cumulative distribution of $M(t)$ i.e.,
\bea
S_0(0,M|t)=\text{Prob}\left[ M(t)\leq M \right].
\eea
We now perform the double inverse Laplace transformation of Eq. \eqref{gen-eq-3} with respect to $k$ and $s$ to get the joint distribution $P_r(M,t_m|t)$ for $t, t_m \gg r^{-1}$ and $t_m \neq t$. The expression of $P_r(M,t_m|t)$ reads
\begin{align}
P_r(M,t_m|t) \simeq r^2 e^{-r t ~\mathcal{U}(M, r)}~\frac{d\bar{S}_0(0,M|r) }{dM}.
\label{gen-eq-4}
\end{align}
Finally to get the distribution of $t_m$, we integrate $P_r(M,t_m|t)$ for all $M$ to yield
\begin{align}
P_r(t_m|t) \simeq \frac{1}{t} \left[ e^{-rt~\mathcal{U}(M \to \infty, r)}-e^{-rt~\mathcal{U}(M \to 0, r)}\right].
\label{gen-eq-5}
\end{align}
Using $\mathcal{U}(M \to \infty, r) =0$ and $\mathcal{U}(M \to 0, r) \geq 0$, we get 
\begin{align}
P_r(t_m|t) \simeq \frac{1}{t}.
\label{gen-eq-6}
\end{align}
In Figure \ref{dist-tm-all}, we have compared our analytic expression for $P_r(t_m|t)$ (given in Eq. \eqref{gen-eq-6}) to the results of the numerical simulations for five different stochastic processes namely diffusion, diffusion with positive and negative drift, random acceleration and Ornstein Uhlenbeck process. It is important to remark that random acceleration process (RAP) is a non-Markov process while the others are Markovian in nature. We observe an excellent agreement between our analytical prediction and the simulations in all the cases (we have consigned the details of simulation in Appendix \ref{simulations}). As stressed before, this result (\ref{gen-eq-6}) is indeed independent of the nature of the underlying stochastic process. 

We end this section by discussing the origin of this universal limiting distribution for $t_m$. Consider a long trajectory $x(\tau)$ with $0\leq \tau \leq t$ with many resetting intervals. However, these intervals are statistically independent since the entire configuration (all the variables) is renewed after each resetting event, and hence there are no correlations between the intervals. Hence, we can map our problem to a simple example of a discrete time intervals $\{\tau_1, \tau_2, \cdots, \tau_{N-1} \}$ of $N-1$ IID entries, each drawn from a PDF $p(\tau)$. As the observation time $t$ becomes large, effect of the last interval $\tau_{N}$ becomes negligible i.e., $t-\tau_N \approx t$. Note that the maximum $M$ can be in any one of these $N$ intervals with equal probability $1/N$ for a given $N$. However, $N$ is a random variable in a fixed time interval $t$. In fact, $N$ is a Poisson process with $P(N)=\frac{(rt)^N}{N!}e^{-rt}$. Taking the average, we find that arg-max $t_m$ converges to a uniform distribution at large $t$ namely \eref{gen-eq-6}. 
Thus, this result is completely universal, i.e., independent of the PDF $p(\tau)$ and the underlying stochastic motion at large time. The essential key point in this derivation is that resetting makes the intervals completely independent to each other. Thus, even though for simple process like Brownian motion (without resetting), computation of $P_0(t_m|t)$ can not be made using this simple argument, resetting simplifies the problem elegantly in many folds and the resulting density pertains to a uniform distribution as we have already shown.

\section{Conclusion}
\label{conclusion}
In summary, we have extensively studied statistics of the maximum distance $M(t)$ and the time $t_m(t)$ taken to reach this maximum distance (upto an observation time $t$) by a stochastic process which is subject to a resetting mechanism. Resetting occurs at a constant rate $r$ which reinstates the particle back to its initial position intermittently. The process is renewed after each resetting event, and the memory from the previous trial is erased. Utilizing this key property, we derive a renewal formula (\eref{main-eq-5}) for the joint distribution of $M(t)$ and $t_m$ in the presence of resetting ($r>0$) in terms of the same but with $r=0$. Our derivation is quite generic and holds for both Markov and non-Markov underlying process. 

Next, we use \eref{main-eq-5} and marginalize it to study statistics of $M(t)$ and $t_m(t)$ respectively. We choose simple diffusion and diffusion with drift as the underlying process and add resetting to them. In the case of simple diffusion, we explicitly computed all moments of $M(t)$ from which we showed that they have logarithmic time dependence at large time. Surprisingly, although the position density converge to time independent steady state, $\langle M^n(t) \rangle$ grows with time but rather slowly with logarithmic dependence (see also \cite{MajumdarMori2020}). Our results are consistent to demonstrate that the limiting distribution of $M(t)$ belongs to a Gumbel class which was recently understood from the EVS theory of weakly interacting random variables \cite{Restart1,review,EVS-review-cor}. For the drift-diffusion case, we have also computed all moments of $M(t)$. As in the case of simple diffusion, we again observe the logarithmic growth for $\langle M^n(t) \rangle$ despite the fact that  drift-diffusion process reaches a steady state in the presence of resetting. Finally, we do a consistency check to show that the distribution of the maximum reproduces the Gumbel law as expected from the EVS theory.

We then turn our attention to the statistics of the arg-max $t_m$. We first derived the moments generating function for $t_m(t)$ in the Laplace space for simple diffusion with reset in Eq. \eqref{BMeq-13}. This allowed us to compute the first two moments exactly for all $t$. We next show that the arg-max density at large time converges to a uniform distribution which only depends on the observation time $t$ but not on the specifics of the underlying process. The sub-leading correction terms are shown to be process dependent. For the drift-diffusion process, computation of the higher order moments beyond the first one becomes quite tedious however at large time we extracted the leading contributions with logarithmic sub-leading terms. The density at large time is again uniform with process dependent correction terms.

Borrowing wisdom from these exact results, we next analyzed the density of $t_m$ for generic stochastic processes subject to stochastic resetting with a rate $r$. Following an asymptotic expression for the joint distribution of $M(t)$ and $t_m(t)$, we show that indeed at large time $P_r(t_m|t) \simeq \frac{1}{t}$ which is independent of the underlying process but the sub-leading terms are naturally process dependent as demonstrated for simple diffusion and drift-diffusion. We provide a probabilistic interpretation of the result based on the renewal property of the resetting phenomena. Numerical simulations covering many of the above-mentioned processes are in perfect agreement with our analytical predictions.

Estimating the probability of extreme events is an important problem in statistics as well as in statistical, mathematical and condensed matter physics and in other interdisciplinary subjects. Despite its paramount relevance, EVS poses serious computational challenges and exact results are scarce. In this paper, we showcase one such example for a renewal process namely stochastic resetting which has gained immense interest in recent times. Although the subject has been very dynamic from the perspective of non-equilibrium transport properties or the first passage estimation, obtaining exact results and limiting distributions related to EVS has been very limited. Finding such results can be often subtle due to the complexity inherent to the systems. However, drawing inspirations from the canonical model systems such as simple diffusion and diffusion with drift, we have been able to unravel some of the universal features of extremals in stochastic resetting systems. We conclude by stating that the scope of the current formalism is not restricted only to instantaneous resetting but also can accommodate scenarios when returns are spatio-temporally correlated \cite{HRS,return-0,return-1,return-2,return-3,return-4}.

\begin{acknowledgements}
Prashant Singh acknowledges useful discussions during ICTS-TIFR programs `BSSP X' and `Fluctuations in Nonequilibrium Systems: Theory and Applications'.
Arnab Pal gratefully acknowledges support
from the Raymond and Beverly Sackler Post-Doctoral Scholarship
and the Ratner Center for Single Molecule Science at
Tel-Aviv University.
\end{acknowledgements}

\appendix
\begin{widetext}
\section{Derivation of the renewal formula in \eref{main-eq-5}}
\label{Joint-LT}
In this section, we present the derivation of the renewal formula in \eref{main-eq-5} which was presented in the main text. For brevity, we recall that the maximum $M$ can occur in any of the $N$-intervals. Following the main text, these contributions are given by
\begin{align}
\mathcal{C}_1=    &\left(\prod _{i=1}^{N}\int_0^\infty d\tau_i\right) \int_0^{\infty}~d\tau~P_0(M,\tau|\tau_1) p(\tau_1)  \left[\prod_{i = 2}^{N-1}~S_0(0,\tau_i|M) p(\tau_i) \right]  \left[ e^{-r\tau_N} S_0(0,\tau_N|M) \right]  ~\delta(t-\sum_{i=1}^N \tau_i)~\delta(\tau-t_m), \nonumber \\
 \mathcal{C}_2 = &  \left(\prod _{i=1}^{N}\int_0^\infty d\tau_i\right)~\int_0^{\infty}~d\tau~P_0(M,\tau|\tau_2) p(\tau_2) \left[ \prod_{i =1, i \neq 2}^{N-1}~S_0(0,\tau_i|M) p(\tau_i)\right]  \left[ e^{-r\tau_N} S_0(0,\tau_N|M) \right]  \delta(t-\sum_{i=1}^N \tau_i)~ \delta(\tau_1+\tau-t_m), \nonumber \\
 \mathcal{C}_3 = & \cdots \nonumber \\
 \mathcal{C}_4 = & \cdots \nonumber \\
 \cdots \nonumber \\
 \mathcal{C}_N= & \left(\prod _{i=1}^{N}\int_0^\infty d\tau_i\right)~\int_0^{\infty}~d\tau~P_0(M,\tau|\tau_N) e^{-r\tau_N}  \left[ \prod_{i =1}^{N-1}~S_0(0,\tau_i|M) p(\tau_i) \right]\delta(t-\sum_{i=1}^N \tau_i)~ \delta(\sum_{i=1}^{N-1}\tau_i+\tau-t_m).
    \label{contr-all-SM}
\end{align}
To obtain joint distribution $P_r \left(M,t_m,N|t\right)$, one has to sum all contributions $\mathcal{C}_1,~\mathcal{C}_2,...,\mathcal{C}_N$. Inserting the contributions from Eqs. \eqref{contr-all-SM} and noting that $e^{-r \tau _N} = \frac{1}{r} p(\tau _N)$ with $p(\tau _i) = r e^{-r \tau _i}$, the joint distribution $\mathcal{P}_r \left(M,t_m,N|t\right)$ can be formally written as  
\begin{align}
\mathcal{P}_r\left(M, t_{m},N|t \right) &= \frac{1}{r}\sum _{j=1}^{N} \int_{0}^{\infty} d \tau _j~  d \tau ~P_0\left(M, \tau|\tau _j \right) p(\tau _j) 
 \left( \prod _{j'=1,j' \neq j}^{N}\int_{0}^{\infty} d \tau _{j'} S_0(0, \tau _{j'}|M) p(\tau _{j'}) d \tau _{j'}\right) \nonumber \\ & \times \delta \left( \sum _{i=1}^{j-1}\tau _i+\tau - t_{max} \right)\delta(t-\sum_{i=1}^N \tau_i),
\label{main-eq-2-SM}
\end{align}
where the second delta function ensures that the total observation time is $t$. To simplify this expression, it is useful to perform Laplace transformations with respect to $t~(\to s)$ and $t_m~(\to k)$. Denoting the Laplace transformation of $\mathcal{P}_r\left(M, t_{m},N|t \right)$ by $\mathcal{Z}_r\left(M, k, N|s \right)$, we take the Laplace transformation of Eq. \eqref{main-eq-2-SM} to yield
\begin{align}
\mathcal{Z}_r\left(M, k, N|s \right) & = \sum _{m=1}^{N}\mathcal{Z}_0\left(M, k|r+s \right) \left[ r \bar{S}_0(0, s+r+k|M)\right]^{m-1}  \left[ r \bar{S}_0(0, s+r|M)\right]^{N-m}, \label{main-eq-3-SM} \\
& = \frac{\mathcal{Z}_0\left(M, k|r+s \right)}{r \left[\bar{S}_0(0, s+r|M)-\bar{S}_0(0, s+r+k|M) \right]} \left[\left( r \bar{S}_0(0, s+r|M)\right)^{N}-\left( r\bar{S}_0(0, s+r+k|M)\right)^{N}  \right].
\label{main-eq-4-SM}
\end{align}
In the first line, we have used the notation $\bar{S}_0(0, s,M)$ as the Laplace transformation of $S_0(0, t|M)$. To get the joint distribution of $M$ and $t_m$ in the Laplace space of $k$ and $s$, we sum $\mathcal{Z}_r\left(M, k, N|s \right)$ for all values of $N$ from 1 to $\infty$ which results in
\begin{align}
\mathcal{Z}_r\left(M, k|s \right) &=\sum _{N=1}^{\infty} \mathcal{Z}_r\left(M, k, N|s \right),\nonumber \\
&= \frac{\mathcal{Z}_0\left(M, k|r+s \right)}{\left[1-r \bar{S}_0(0, s+r|M)\right]~\left[1-r\bar{S}_0(0, s+r+k|M) \right]},
\label{main-eq-5-SM}
\end{align}
where in going to the second line from the first line, we have substituted $ \mathcal{Z}_r\left(M, k, N|s \right)$ from Eq. \eqref{main-eq-4-SM}. This concludes the proof for \eref{main-eq-5} in the main text.
\end{widetext}

\section{Asymptotic forms of $\langle M^n(t) \rangle$ for Brownian motion}
 \label{asy-sca-M-BM}
 In this appendix, we derive the asymptotic behaviour of $\langle M^n(t) \rangle$ for Brownian motion whose exact expression is given in Eq. \eqref{BMeq-6}. Below, we look at the behaviour of $\langle M^n(t) \rangle$ for large and short times separately.
 \subsection{Case I: $\langle M^n(t) \rangle$ for $t \gg \frac{1}{r}$}
 To begin with, we consider the Laplace transformation of $P_r(M|t)$ from Eq. \eqref{dist-max-BM-eq-1} and rewrite here as
 \begin{align}
\mathcal{Z}_r\left(M, k=0|s \right) = \frac{\sqrt{2} \left( r+s\right)^{3/2} ~e^{-\sqrt{2(r+s)}M}}{ \left( s+r e^{-\sqrt{2(r+s)}M}\right)^2}.
\label{appen-dist-max-BM-eq-1}
\end{align}
For small $r \gg s$ (which corresponds to large $r \ll t^{-1} $), we approximate $r+s \simeq s$ in Eq. \eqref{appen-dist-max-BM-eq-1} to yield
 \begin{align}
\mathcal{Z}_r\left(M, k=0|s \right) \simeq  \frac{\sqrt{2} r^{3/2} ~e^{-\sqrt{2r}M}}{ \left( s+r e^{-\sqrt{2 r}M}\right)^2}.
\label{appen-dist-max-BM-eq-2}
\end{align}
To get the distribution in the time domain, we perform the inverse Laplace transform of Eq. \eqref{appen-dist-max-BM-eq-2} by using $\mathcal{L}_{s \to t} ^{-1} \left[ \frac{1}{(s+a)^2}\right] = t e^{-a t}$ for $a \geq 0$ to yield
\begin{align}
P_r(M|t) \simeq \sqrt{2r^3} t e^{-\sqrt{2r}M} \text{exp} \left(- r t e^{-\sqrt{2r}M}  \right).
\label{appen-dist-max-BM-eq-3} 
\end{align}
We emphasise that the approximate equality in this equation indicates that it is valid only for $t \gg \frac{1}{r}$. We use this form of the distribution to get the moments at large $t$ as
\begin{align}
\langle M^n(t) \rangle &= \int _0 ^{\infty} dM M^n~P_r(M|t), \label{appen-dist-max-BM-eq-44}\\
& \simeq \frac{1}{(2r)^{n/2}} \log ^{n}(rt) + O\left( \log ^{n-1}(rt)\right),
\label{appen-dist-max-BM-eq-4}
\end{align}
where in going to the second line, we have inserted $P_r(M|t)$ from Eq. \eqref{appen-dist-max-BM-eq-3} and performed the integration over $M$. Comparing Eq. \eqref{appen-dist-max-BM-eq-4} with the scaling form in Eq. \eqref{BMeq-6}, we find that the scaling function $H_n(z)$ at large $z$ is given by
\begin{align}
H _n (z) \simeq \log ^n z + O\left( \log ^{n-1} z\right).
\label{appen-dist-max-BM-eq-5}
\end{align}
 This result has been quoted in Eq. \eqref{BMeq-8} of the main text.

\subsection{Case II: $\langle M^n(t) \rangle$ for $t \ll \frac{1}{r}$}
To find the moments at times $t \ll \frac{1}{r}$, we need the form of the distribution $P_r(M|t)$ at these times as indicated by Eq. \eqref{appen-dist-max-BM-eq-44}. In the Laplace space, this will correspond to the $r \ll s$ behaviour of $\mathcal{Z}_r\left(M, k=0|s \right)$ in Eq. \eqref{appen-dist-max-BM-eq-1}. By direct expansion of Eq. \eqref{appen-dist-max-BM-eq-1} in $r$, we find 
 \begin{align}
\mathcal{Z}_r\left(M, k=0|s \right) &\simeq \sqrt{\frac{2}{s}} e^{-\sqrt{2s}M} + \frac{r}{2 s^{3/2}} \left[ \left( 3 \sqrt{2}-2 M  \sqrt{s}\right)\right. \nonumber \\
& \left. \times e^{-\sqrt{2 s} M} - 4\sqrt{2} ~ e^{-2\sqrt{2 s} M}\right].
\label{appen-dist-max-BM-eq-6}
\end{align}
We next use Eq.\eqref{BMeq-5} (first line) to get the moments in the Laplace space and then perform inverse Laplace inversion to yield
\begin{align}
\langle M^n(t) \rangle \simeq \frac{n!}{2 ^{\frac{n}{2}}\Gamma \left(1+\frac{n}{2} \right)}~t^{\frac{n}{2}}+\frac{\mathcal{B}_n r}{2^{n/2}} t^{\frac{n}{2}+1},~~\text{with}
\label{appen-dist-max-BM-eq-7}
\end{align}  
\begin{align}
\mathcal{B}_n = \frac{\Gamma (n+1) \left( 3 \sqrt{2}-2^{\frac{3}{2}-n}\right)-\sqrt{2}\Gamma (n+2)}{2 \sqrt{2}~ \Gamma \left( \frac{n+4}{2}\right)}.
\label{Bn-exp}
\end{align} 
Finally comparing this equation with the scaling form in Eq. \eqref{BMeq-6}, we find that the scaling function $H_n(z)$ at small $z$ is given by
\begin{align}
H _n (z) \simeq \frac{n!}{\Gamma \left(1+\frac{n}{2} \right)} z^{\frac{n}{2}} + \mathcal{B}_n z^{\frac{n}{2}+1},
\label{appen-dist-max-BM-eq-8}
\end{align}
which is Eq. \eqref{BMeq-9} in the main text.

\section{Derivation of $\langle t_{m}(t) \rangle$ and $\langle t^2_{m}(t) \rangle$ for Brownian motion}
\label{appen-tm-BM}
In this appendix, we derive the exact expressions of $\langle t_{m}(t) \rangle$ and $\langle t^2_{m}(t) \rangle$ which are presented in Eqs. \eqref{BMeq-18} and \eqref{BMeq-20} respectively. To derive these results, we insert the expression of $\mathcal{Z}_{r} \left(k|s \right)$ from Eq. \eqref{BMeq-13} in Eq. \eqref{mom-tm-eq-46} for $n=1,2$ to get
\begin{align}
\int _{0}^{\infty} dt e^{-st} \langle t_{m}(t) \rangle &= \frac{2 r^2+3 r s+s^2}{4 s^2 (r+s)^2} + \frac{ \log \left( \frac{r+s}{s}\right)}{4 r s}, \label{appen-tm-BM-eq-1}\\
\int _{0}^{\infty} dt e^{-st} \langle t_{m}^2(t) \rangle &= \frac{16 r^2+36 r s+15 s^2}{24 s^3 (r+s)^2} -\frac{\text{Li}_2\left( -\frac{r}{s}\right)}{12 r s (r+s)} \nonumber \\
&+\frac{(4 r+s)\log \left( \frac{r+s}{s}\right)}{24 r s^2 (r+s)}.
\label{appen-tm-BM-eq-2}
\end{align}
Let us invert Eq. \eqref{appen-tm-BM-eq-1} with respect to $s$. To this aim, we use the following inverse Laplace transformations:
\begin{align}
&\mathcal{L}_{s \to t}^{-1}\left[\frac{\log \left( \frac{r+s}{s}\right)}{s} \right]=\gamma _E + \Gamma(0, r t) + \log r t, \label{ILT-4} \\
&\mathcal{L}_{s \to t}^{-1}\left[ \frac{2 r^2+3 r s+s^2}{4 s^2 (r+s)^2} \right] = \frac{2 r t+e^{-r t}-1}{4r}. \label{ILT-5}
\end{align}
Using these two inverse Laplace transforms in Eq. \eqref{appen-tm-BM-eq-1}, we recover the expression of $\langle t_{m}(t) \rangle$ as written in Eq. \eqref{BMeq-18}. We next look at Eq. \eqref{appen-tm-BM-eq-2} to get $\langle t_m^2(t) \rangle$ for which we need three inverse Laplace transformations. Inverse Laplace transform of the first term in the RHS of Eq. \eqref{appen-tm-BM-eq-2} is given by
\begin{align}
\mathcal{L}_{s \to t}^{-1}\left[\frac{16 r^2+36 r s+15 s^2}{24 s^3 (r+s)^2} \right]=&\frac{1}{24 r^2}\left[ e^{-r t}(9+5 r t)-9 \right. \nonumber \\
& \left.+4 r t+8 r^2 t^2\right].
\label{ILT-6} 
\end{align}
Next, we turn to the second term in the RHS of Eq. \eqref{appen-tm-BM-eq-2}. Note that this term is the product of $\frac{1}{r+s}$ and $\frac{1}{s}\text{Li}_2 \left(-\frac{r}{s} \right)$ which implies that the convolution theorem for Laplace transforms can be directly used. Doing so, we find
\begin{align}
&\mathcal{L}_{s \to t}^{-1}\left[ \frac{\text{Li}_2 \left(-\frac{r}{s} \right)}{s(r+s)}\right] = \int _{0}^{t} d t_1 e^{-r(t-t_1)} f_1(t_1),~~~\text{where} \label{ILT-7} \\
&f_1(t_1) = \mathcal{L}_{s \to t_1}^{-1}\left[ \frac{\text{Li}_2 \left(-\frac{r}{s} \right)}{s}\right], \\
&~~~~~~~= - r t_1 ~_3F_3 \Big(\{1,1,1 \},\{2,2,2 \},\{-r t_1 \} \Big).\label{ILT-8} 
\end{align}
We now look at the third term in the R.H.S. of Eq. \eqref{appen-tm-BM-eq-2}. Once again, we notice that it has the product form because of which we use the convolution theorem.  The inverse Laplace transform then reads 
\begin{align}
\mathcal{L}_{s \to t}^{-1}\left[\frac{(4 r+s)\log \left( \frac{r+s}{s}\right)}{24 r s^2 (r+s)} \right]= \int _{0}^{t} d t_1 f_2(t-t_1) f_3(t_1),
\label{ILT-8}
\end{align}
where the functions $f_{2}(t)$ and $f_{3}(t)$ are given by
\begin{align}
f_2(t) &= \mathcal{L}_{s \to t}^{-1} \left[\log \left( \frac{r+s}{s}\right)\right] , \\
&=\frac{1-e^{-r t}}{t},\\
f_3(t) & = \mathcal{L}_{s \to t}^{-1} \left[\frac{(4 r+s)}{24 r s^2 (r+s)}\right], \\
& = \frac{4 r t-3+3 e^{-r t}}{24 r^2}.
\end{align}
Finally inserting Eqs. \eqref{ILT-6}, \eqref{ILT-7} and \eqref{ILT-8} in Eq. \eqref{appen-tm-BM-eq-2}, we recover the form of $\langle t_{m}^2(t) \rangle$ as quoted in Eq. \eqref{BMeq-20}.

\section{Asymptotic forms of $\langle t_m^n(t) \rangle$ for Brownian motion}
\label{appen-asy-tm-BM}
This appendix deals with the derivation of the forms of $\langle t_m^n(t) \rangle$  for Brownian motion when $t \ll r^{-1}$ and $t \gg r^{-1}$. These asymptotic forms are presented in Eqs. \eqref{BMeq-16-new} and \eqref{BMeq-16}. To begin with, we analyse $\mathcal{Z}_{r} \left(k|s \right)$ in Eq. \eqref{BMeq-13} in various limits of $s r^{-1}$ using which we compute the distribution $P_r(t_m|t)$ for different regimes of $t$. We then use $P_r(t_m|t)$ to compute moments. Below we consider the cases $t \ll r^{-1}$ and $t \gg r^{-1}$ separately.
\subsection{Case I: $t \ll \frac{1}{r}$} 
In the Laplace space of $s$, the limit $t \ll \frac{1}{r}$ corresponds to $r s^{-1} \to 0$ which for fixed $s$ means $r \to 0$. Taking the limit $r \to 0$ in $\mathcal{Z}_{r} \left(k|s \right)$ in Eq. \eqref{BMeq-13}, we get
\begin{align}
\mathcal{Z}_{r} \left(k|s \right) \simeq \frac{1}{\sqrt{s(s+k)}} + \frac{r}{s^{3/2}} \left[ \frac{1}{2 \sqrt{s+k}} -\frac{1}{\sqrt{s}+\sqrt{s+k}} \right].
\label{appen-asy-tm-BM-eq-1}
\end{align}
We next invert this Laplace transform to obtain the distribution $P_r(t_m|t)$ using which we compute moments. Note that we have to perform double inverse Laplace transformations: one from $s \to t$ and the other from $k \to t_m$. To invert Eq. \eqref{appen-asy-tm-BM-eq-1}, we use the following double inverse Laplace transformations:
\begin{align}
&\mathcal{L}_{s \to t}^{-1}\mathcal{L}_{k \to t_m}^{-1} \left[ \frac{1}{\sqrt{s(s+k)}}\right] = \frac{1}{\pi \sqrt{t_m(t-t_m)}}, \label{ILT-1}\\
&\mathcal{L}_{s \to t}^{-1}\mathcal{L}_{k \to t_m}^{-1} \left[ \frac{s^{-3/2}}{(\sqrt{s}+\sqrt{s+k})}\right]= \frac{2}{\pi} \left[ \sqrt{\frac{t-t _m}{t_ m}}\right.\nonumber\\
&~~~~~~~~~~~~~~~~~~~~~~~~~~~~~~~~~~~~~~~\left.- \cos ^{-1}\sqrt{\frac{t_m}{t}}\right].\label{ILT-2}
\end{align}
Using these two equations in $\mathcal{Z}_{r} \left(k|s \right)$ in Eq. \eqref{appen-asy-tm-BM-eq-1}, we get
\begin{align}
P_r(t_m|t) \simeq \frac{1}{\pi \sqrt{t_m(t-t_m)}} +\frac{r}{\pi} \left( 2\cos ^{-1}\sqrt{\frac{t_m}{t}}-\sqrt{\frac{t-t_m}{t_m}}\right).
\label{appen-asy-tm-BM-eq-2}
\end{align}
We emphasize that this expression is valid only  in the limit $r t \to 0$. Finally, we use this form of $P_r(t_m|t)$ to compute the moments of $t_m$ which then read
\begin{align}
\langle t_{m}^n (t) \rangle 
 &\simeq \frac{(2n-1)! t^{n}}{n! (n-1)! 2^{2n-1}}  + \mathcal{A}_n\gamma t^{n+1},~~~\text{with} \label{appen-asy-tm-BM-eq-3-1}\\
  \mathcal{A}_n &= \frac{1}{\Gamma (n+2)} \left[ \frac{\Gamma \left( 3/2+n\right)}{\sqrt{\pi} (n+1)}-\frac{(2n-1)!}{2^{2n} (n-1)!}\right],
  \label{An-exp}
\end{align}
which is Eq. \eqref{BMeq-16-new} in the main text.

\subsection{Case II: $t \gg \frac{1}{r}$} 
We next look at the moments when $t \gg \frac{1}{r}$. Once again we begin with the expression of $\mathcal{Z}_{r} \left(k|s \right)$ in Eq. \eqref{BMeq-13} and analyse it in the limit of large $r$. For large $r$, we approximate $r+s \simeq r$ and $w^{\sqrt{\frac{r+s}{r+s+k}}} \simeq w-\frac{w k  \log w}{2 r}$ and insert them in Eq. \eqref{BMeq-13} to yield
\begin{align}
\mathcal{Z}_{r} \left(k|s \right) \simeq r \int _{0}^{1} \frac{dw}{(s+k+rw)\left( s+rw-\frac{w k \log w}{2}\right)}.
\label{appen-asy-tm-BM-eq-3}
\end{align}
To get the distribution in the time domain from Eq. \eqref{appen-asy-tm-BM-eq-3}, we now use the following standard inverse Laplace transformation
\begin{align}
\mathcal{L}_{s \to t} ^{-1} \left[\frac{e^{-b s}}{s+a} \right] = e ^{-a(t-b)} \Theta(t-b),~~\text{with }a,b \geq 0,
\label{ILT-3}
\end{align} 
in Eq. \eqref{appen-asy-tm-BM-eq-3} to get
\begin{align}
P_r(t_m|t) \simeq 2r \int _{0}^{1} dw\frac{e ^{-rtw}}{2+w \log w} \left[1- \Theta \left(t+\frac{2 t_m}{w \log w} \right) \right].
\label{appen-asy-tm-BM-eq-4}
\end{align}
We are now left with the integration over $w$. This can be done by making the transformation $rtw = y$ in Eq. \eqref{appen-asy-tm-BM-eq-4}. To proceed further, we approximate $\log w = \log y-\log rt \simeq - \log rt$ and $\frac{r t}{\log rt} \to \infty$ for $rt \to \infty$. With these approximations, the integration in Eq. \eqref{appen-asy-tm-BM-eq-4} can be performed explicitly to get
\begin{align}
P_r(t_m|t) \simeq \frac{1}{t} + \frac{\log rt}{2 r t^2}.
\label{appen-asy-tm-BM-eq-5}
\end{align}
Using this form of $P_r(t_m|t) $ for $t \gg r^{-1}$, it is straightforward to show the moment is given by
\begin{align}
\langle t_{m}^n (t) \rangle \simeq \frac{t^n}{n+1} + \frac{t^{n-1} \log (rt)}{2r(n+1)}, \label{appen-asy-tm-BM-eq-6}
\end{align}
which has been mentioned in Eq. \eqref{BMeq-16}.

\section{Derivation of $P_r(t_m|t)$ in Eq. \eqref{BMeq-15-new} for Brownian motion}
\label{appen-dist-BM}
In this appendix, we perform the double inverse Laplace transformation of $\mathcal{Z}_{r} \left(k|s \right)$ in Eq. \eqref{dist_new_eqq_573} to get the distribution $P_r(t_{m}|t)$ in Eq. \eqref{BMeq-15-new}. We first perform the inversion of Eq. \eqref{dist_new_eqq_573} with respect to $k$ for which we use the following:
\begin{align}
\mathcal{L}_{k \to t_m}^{-1} \left[ \frac{\sqrt{k+b}}{k+a}\right] =& \frac{e^{-b t_m}}{\sqrt{\pi t_m}} + \sqrt{b-a} ~e^{-a t_m}\nonumber \\
 & \times \text{Erf}\left( \sqrt{(b-a)t_m}\right),
\label{appen-dist-BM_eq_1222}
\end{align}
with $a, b \geq 0$. Using this in Eq. \eqref{dist_new_eqq_573} by reading appropriately $a$ and $b$, we get
\begin{align}
&\bar{P}_r(t_m|s) = \bar{Q}_1(t_m|s)+\bar{Q}_2(t_m|s), \label{appen-dist-BM_eq_1}
\end{align}
where $\bar{P}_r(t_m|s)$ stands for the Laplace transformation of $P_r(t_m|t)$. Also, the functions $\bar{Q}_1(t_m|s)$ and $\bar{Q}_2(t_m|s)$ are given by
\begin{align}
\bar{Q}_1(t_m|s) &=\int _0^{1} dw \frac{\sqrt{r+s}~e^{-(r+s)t_m}}{\sqrt{\pi t_m}(s+rw)}, \label{appen-dist-BM_eq_2}\\
\bar{Q}_2(t_m|s) &=\int _0^{1} dw \frac{\sqrt{r(r+s)(1-w)}~e^{-(s+rw)t_m}}{s+rw} \nonumber \\
& \times \text{Erf}\left( \sqrt{r t_m(1-w)}\right).
\label{appen-dist-BM_eq_3}
\end{align}
To obtain the distribution $P_r(t_m|t)$, we have to perform inverse Laplace transformation in Eq. \eqref{appen-dist-BM_eq_1}. Let us first perform the inversion for $\bar{Q}_1(t_m|s)$ in Eq. \eqref{appen-dist-BM_eq_2} for which we use the inverse Laplace transformation in Eq. \eqref{appen-dist-BM_eq_1222} with $k$ replaced by $s$. Denoting the inverse Laplace transformation of $\bar{Q}_1(t_m|s)$ by $Q_1(t_m|t)$, we get
\begin{align}
Q_1(t_m|t)& = \frac{e^{-r t}}{\pi \sqrt{t_m(t-t_m)}}+\frac{e^{-r t_m}}{\sqrt{\pi t_m}} \int _{0}^{1} dw \sqrt{r(1-w)} \nonumber\\
& \times e^{-rw(t-t_m)} ~\text{Erf}\left(\sqrt{r(1-w)(t-t_m)} \right).
\label{appen-dist-BM_eq_3}
\end{align}
To perform the integration in the second term, we make the following change in variable: $y = r(1-w)(t-t_m)$, which, in turn, yields
\begin{align}
Q_1(t_m|t)& = \frac{e^{-r t}}{\pi \sqrt{t_m(t-t_m)}} \left[ 1+\frac{\sqrt{\pi}}{r(t-t_m)} \right.\nonumber \\
&\left. \times \int _0^{r(t-t_m)} dy \sqrt{y}~ e^{y} ~\text{Erf}(\sqrt{y})\right].
\label{appen-dist-BM_eq_4}
\end{align}
The integration over $y$ in the second line can be explicitly performed using Mathematica in terms of the generalized hypergeometric functions. The final result reads
\begin{align}
Q_1(t_m|t) = \frac{e^{-r t}~_2F_2 \left(  \{1,1 \},\{1/2, 2 \}, r(t-t_m)\right)}{\pi \sqrt{t_m(t-t_m)}} .
\label{appen-dist-BM_eq_5}
\end{align}
We next turn to $\bar{Q}_2(t_m|s)$ in Eq. \eqref{appen-dist-BM_eq_1}. For $\bar{Q}_2(t_m|s)$, it turns out that one has to follow similar steps as done for $\bar{Q}_1(t_m|s)$. To avoid repetition, we only present here the final expression of $Q_2(t_m|t)$ which reads
\begin{align}
Q_2(t_m|t) &=e^{-r t}\frac{\left[~_2F_2 \left(  \{1,1 \},\{1/2, 2 \}, r(t-t_m)\right)-1\right]}{\pi \sqrt{t_m(t-t_m)}}\nonumber \\
&+ r e^{-r t} \int _0^{1} ~dw ~w ~e^{r t w}  ~ \text{Erf} \left(\sqrt{r w t_{m}} \right) \nonumber \\
&~~~~~~~~~\times  \text{Erf} \left(\sqrt{r w (t-t_{m})} \right).
\label{appen-dist-BM_eq_6}
\end{align}
Finally using Eqs. \eqref{appen-dist-BM_eq_5} and \eqref{appen-dist-BM_eq_6} in Eq. \eqref{appen-dist-BM_eq_1}, we recover the result for $P_r(t_m|t)$ in Eq. \eqref{BMeq-15-new}.

\section{Derivation of $\langle M^n(t) \rangle$ for drift-diffusion process}
\label{M-mom-ddiff}
This appendix provides the derivation of the scaling relation in Eq. \eqref{mom-M-dd} for $\langle M^n(t) \rangle$. We begin with the Laplace transform of $\langle M^n(t) \rangle$ in Eq. \eqref{BMDeq-5}. Looking at this equation, we find that the Laplace transform is product to two terms which leads us to use convolution property for Laplace transforms. Using this property, we get
\begin{align}
\langle M^n(t) \rangle = -\frac{n!}{2^{n/2}} \sum _{k=1}^{\infty}\frac{(-r)^k}{k^n} \int _{0}^{t} d \tau f_4(t-\tau), f_3^n(\tau)
\label{M-mom-ddiff-eq-1}
\end{align}
where the functions $f_3(t)$ and $f_4(t)$ are given below.
\begin{align}
f_3(t) &= \mathcal{L}_{s \to t}^{-1} \left[\frac{1}{\left( \sqrt{s-\beta}+\gamma \right)^n} \right], \nonumber \\
& = \frac{(-1)^{n-1}}{(n-1)!} \left[\frac{e^{\beta t}~\delta _{n,1}}{ \sqrt{\pi t}} -\frac{d^{n-1}}{d \gamma ^{n-1}} \left\{ \gamma e^{(\beta+\gamma^2)t} \text{Erfc}\left( \gamma \sqrt{t}\right)\right\}\right] \nonumber \\
f_4(t)&=\mathcal{L}_{s \to t}^{-1} \left[\frac{s+r}{s^{k+1}} \right], \nonumber \\
& = \frac{t^{k+1}}{\Gamma (k)} + \frac{r t^k}{\Gamma (k+1)},
\end{align} 
where $\beta =-\left(r +\gamma ^2 \right)$ and $\gamma = -v/\sqrt{2}$. Inserting the forms of $f_3(t)$ and $f_4(t)$ in the expression of $\langle M^n(t) \rangle$ in Eq. \eqref{M-mom-ddiff-eq-1} and then changing the variable $\tau = tw$, we get the scaling relation in Eq. \eqref{mom-M-dd}. \\

\begin{widetext}
\section{Derivation of $P_r(t_m|t)$ in \eref{density-arg-max-DD} for drift-diffusion process}
\label{Appendix-drift-diffusion-tmax-PDF}
In this section, we present derivation for the density $P_r(t_m|t)$ namely \eref{density-arg-max-DD} which was presented in the main text. We recall from the main text (\eref{BMDeq-14-v}) that 
\begin{align}
\mathcal{Z}_{r} \left(k|s \right)  &\simeq \frac{\sqrt{v^2+2(s+r)}-v}{\sqrt{v^2+2(s+k+r)}-v}~ \int _{0}^{1}~dw~ \frac{(r+s+k)}{\left(s+r w \right) \left( s+k+r w\right)}.
\label{BMDeq-14-v-SM}
\end{align}
To invert this expression, we do the inverse Laplace transforms with respect to $k$ and  $s$ respectively. Next, we perform the integral. Following the first inversion, we rewrite
\begin{align}
   \mathcal{Z}_{r} \left(t_m|s \right)  \simeq \left[ \sqrt{v^2+2(s+r)}-v   \right] \int _{0}^{1}~\frac{dw}{\left(s+r w \right)} \mathcal{L}_{k \to t_m}^{-1}
   \left[ \underbrace{ \frac{s+r+k}{s+rw+k} \frac{1}{\sqrt{v^2+2(s+k+r)}-v}}_{\mathcal{I}(k)} \right].
\end{align}
We can now rearrange $\mathcal{I}(k)$ and then do the inversion to find
\begin{align}
    \mathcal{L}_{k \to t_m}^{-1}
       \left[ \mathcal{I}(k) \right]&=  \mathcal{L}_{k \to t_m}^{-1}
   \left[\frac{1}{2}\frac{v}{s+rw+k}+\frac{1}{2}\frac{\sqrt{v^2+2(s+k+r)}}{s+rw+k} \right] \nonumber \\
   &=\frac{e^{-(\frac{v^2}{2}+s+r)t_m}}{\sqrt{2\pi t_m}}+\frac{v e^{-(s+rw)t_m}}{2}+\frac{\sqrt{\frac{v^2}{2}+r(1-w)}}{\sqrt{2}}e^{-(s+rw)t_m} \text{Erf}\left[ \sqrt{\left(\frac{v^2}{2}+r(1-w) \right)t_m} \right].
\end{align}
Now, we need to perform the Laplace inversion wrt $s$ i.e.,
\begin{equation}
    P_r(t_m|t)=\mathcal{L}_{s \to t}^{-1}\left[  \mathcal{Z}_{r} \left(t_m|s \right)  \right],
    \label{pr-inversion}
\end{equation}
where we have
\begin{align}
  \mathcal{Z}_{r} \left(t_m|s \right)&= \int_0^1~dw~ \frac{e^{-(\frac{v^2}{2}+s+r)t_m}}{\sqrt{2\pi t_m}} \frac{\sqrt{v^2+2(s+r)}-v}{s+rw} \nonumber \\
  &+\int_0^1~dw~\frac{v e^{-(s+rw)t_m}}{2}\frac{\sqrt{v^2+2(s+r)}-v}{s+rw} \nonumber \\
  &+\int_0^1~dw~\frac{\sqrt{\frac{v^2}{2}+r(1-w)}}{\sqrt{2}}~\text{Erf}\left[ \sqrt{\left(\frac{v^2}{2}+r(1-w) \right)t_m} \right]~e^{-(s+rw)t_m}\frac{\sqrt{v^2+2(s+r)}-v}{s+rw}.
  \label{zr-app}
\end{align}
Substituting \eref{zr-app} into \eref{pr-inversion} and performing the Laplace inversions, we obtain
\begin{align}
    P_r(t_m|t) &\simeq \frac{e^{-(\frac{v^2}{2}+r)t_m}}{\sqrt{2\pi t_m}}\int_0^1~dw~ \mathcal{J}_\ell(t,w)+
    \frac{v }{2}\int_0^1~dw~e^{-rwt_m}~ \mathcal{J}_\ell(t,w) \nonumber \\
&    +\int_0^1~dw~\frac{\sqrt{\frac{v^2}{2}+r(1-w)}}{\sqrt{2}}~\text{Erf}\left[ \sqrt{\left(\frac{v^2}{2}+r(1-w) \right)t_m} \right]~e^{-rwt_m} ~ \mathcal{J}_\ell(t,w), \nonumber \\
&\simeq \mathbb{I}_1 (t_m,t)+\mathbb{I}_2(t_m,t)+\mathbb{I}_3(t_m,t),
\label{Pr-DD-all-1}
\end{align}
where we have used the following inverse Laplace transform to arrive at \eref{Pr-DD-all-1}
\begin{align}
 \mathcal{J}_\ell(t,w)&=   \mathcal{L}_{s \to t}^{-1} \left[  \frac{\sqrt{v^2+2(s+r)}-v}{s+rw} e^{-st_m} \right]=-ve^{-rw(t-t_m)} \Theta(t-t_m)\nonumber \\
    &+\sqrt{2} \left[ \frac{e^{-(t-t_m)(\frac{v^2}{2}+r)}}{\sqrt{\pi (t-t_m)}}+e^{-rw(t-t_m)}\sqrt{\frac{v^2}{2}+r(1-w)}~ \text{Erf}\left( \sqrt{(t-t_m)(\frac{v^2}{2}+r-rw)} \right)  \right] 
    \Theta(t-t_m).
\end{align}
The $\mathbb{I}$-functions introduced in \eref{Pr-DD-all-1} are formally defined in the following. They can also be simplified occasionally. The first component in \eref{Pr-DD-all-1} reads
\begin{align}
    \mathbb{I}_1(t_m,t)&=\frac{e^{-(\frac{v^2}{2}+r)t_m}}{\sqrt{2\pi t_m}}\int_0^1~dw~ \mathcal{J}_\ell(t,w), \nonumber \\
    &=\frac{e^{-(\frac{v^2}{2}+r)t}}{\pi \sqrt{t_m(t-t_m)}}
    -\frac{ve^{-(\frac{v^2}{2}+r)t_m}}{ \sqrt{2\pi t_m}}~ \frac{1-e^{-r(t-t_m)}}{r(t-t_m)}
    +\frac{e^{-(\frac{v^2}{2}+r)t}}{\pi \sqrt{t_m(t-t_m)}} \int_{\frac{v^2}{2r}}^{\frac{v^2}{2r}+1}~dz~e^{r(t-t_m)z} \sqrt{rz}~\text{Erf}\left[ \sqrt{r(t- t_m)} \sqrt{z}  \right], 
    \nonumber \\
    &=\frac{e^{-(\frac{v^2}{2}+r)t}}{\pi \sqrt{t_m(t-t_m)}}
    -v\frac{e^{-(\frac{v^2}{2}+r)t_m}}{ \sqrt{2\pi t_m}}~ \frac{1-e^{-r(t-t_m)}}{r(t-t_m)}
    \nonumber \\
    &+\frac{e^{-(\frac{v^2}{2}+r)t}}{\pi (t-t_m) \sqrt{\pi t_m}}\left( -\frac{v^2 \, _2F_2\left[1,1;\frac{1}{2},2;\frac{(t-t_m) v^2}{2}\right]}{2 r}+\left(\frac{v^2}{2 r}+1\right) \, _2F_2\left[1,1;\frac{1}{2},2;r (t-t_m) \left(\frac{v^2}{2 r}+1\right)\right]-1 \right).
    \label{def-I1}
\end{align}
Similarly, the second component in \eref{Pr-DD-all-1} gives
\begin{align}
    \mathbb{I}_2(t_m,t)&= \frac{v }{2}\int_0^1~dw~e^{-rwt_m}~ \mathcal{J}_\ell(t,w) ,\nonumber \\
    &=\frac{v e^{-(\frac{v^2}{2}+r)(t-t_m)}}{\ \sqrt{2\pi (t-t_m)}}~\frac{1-e^{-rt_m}}{rt_m}-\frac{v^2}{2}\frac{1-e^{-rt}}{rt}+\frac{ve^{-(\frac{v^2}{2}+r)t}}{\sqrt{2}}  \int_{\frac{v^2}{2r}}^{\frac{v^2}{2r}+1}~dz~e^{rtz} \sqrt{rz}~\text{Erf}\left[ \sqrt{r(t-t_m)} \sqrt{z}  \right].
    \label{def-I2}
\end{align}
Finally, simplifying the third component in \eref{Pr-DD-all-1}, we get
\begin{align}
    \mathbb{I}_3(t_m,t)&=\int_0^1~dw~\frac{\sqrt{\frac{v^2}{2}+r(1-w)}}{\sqrt{2}}~\text{Erf}\left[ \sqrt{\left(\frac{v^2}{2}+r(1-w) \right)t_m} \right]~e^{-rwt_m} ~ \mathcal{J}_\ell(t,w), \nonumber \\
    &=\frac{e^{-(\frac{v^2}{2}+r)t}}{\sqrt{\pi(t-t_m)}}  \int_{\frac{v^2}{2r}}^{\frac{v^2}{2r}+1}~dz~e^{rt_m z} \sqrt{rz}~\text{Erf}\left[ \sqrt{r t_m} \sqrt{z}  \right]
    -\frac{ve^{-(\frac{v^2}{2}+r)t}}{\sqrt{2}}  \int_{\frac{v^2}{2r}}^{\frac{v^2}{2r}+1}~dz~e^{rtz} \sqrt{rz}~\text{Erf}\left[ \sqrt{r t_m} \sqrt{z}  \right],\nonumber \\
    &+r e^{-(\frac{v^2}{2}+r)t} \int_{\frac{v^2}{2r}}^{\frac{v^2}{2r}+1}~dz~z~e^{rtz}~ \text{Erf}\left[ \sqrt{r t_m} \sqrt{z}  \right]\text{Erf}\left[ \sqrt{r (t-t_m)} \sqrt{z}  \right]     
    \nonumber \\
    &=\frac{e^{-(\frac{v^2}{2}+r)t}}{\pi t_m \sqrt{\pi(t-t_m)}} \left( -\frac{v^2 \, _2F_2\left[1,1;\frac{1}{2},2;\frac{t_m v^2}{2}\right]}{2 r}+\left(\frac{v^2}{2 r}+1\right) \, _2F_2\left[1,1;\frac{1}{2},2;r t_m \left(\frac{v^2}{2 r}+1\right)\right]-1 \right) \nonumber \\
    &-\frac{ve^{-(\frac{v^2}{2}+r)t}}{\sqrt{2}}  \int_{\frac{v^2}{2r}}^{\frac{v^2}{2r}+1}~dz~e^{rtz} \sqrt{rz}~\text{Erf}\left[ \sqrt{r t_m} \sqrt{z}  \right]\nonumber \\
    &+r e^{-(\frac{v^2}{2}+r)t} \int_{\frac{v^2}{2r}}^{\frac{v^2}{2r}+1}~dz~z~e^{rtz}~ \text{Erf}\left[ \sqrt{r t_m} \sqrt{z}  \right]\text{Erf}\left[ \sqrt{r (t-t_m)} \sqrt{z}  \right].
    \label{def-I3}
\end{align}
Joining all the $\mathbb{I}$-functions results in the expression (\ref{density-arg-max-DD}) for $P_r(t_m|t)$ which was announced in the main text.

\section{Details of the model systems used in the simulation in \sref{gen-process}}
\label{simulations}
In this section, we present details of the processes used in the simulation in \sref{gen-process}. We have used four different model systems as underlying processes. Three of them are Markovian in nature while one is a non-Markov process. All of them are subjected to resetting at a rate $r$ which means that the time intervals between the resetting events were taken from an exponential distribution namely $p(\tau)=re^{-r\tau}$. We observe trajectories governed by these processes for a fixed observation time $t$, and compute the maximum displacement $M$ that it undertook by this time. Moreover, we also note down the time $t_m$ at which this maximum took place. Details of the model systems are as follows:
\begin{enumerate}
    \item \textbf{Simple diffusion:} Motion of the particle for a simple diffusing particle is given by 
    \bea
    \frac{dx}{d\tau}=\eta(\tau),
    \eea
    where $\eta(\tau)$ is the Gaussian white noise with mean zero and variance $2D\tau$. This is a Markov process.
    \item \textbf{Diffusion with drift:} Here, we consider the diffusing particle in the presence of a drift velocity $v>0$ so that
    \bea
    \frac{dx}{d\tau}=v+\eta(\tau),
    \eea
    which is also a Markov process.
    \item \textbf{Random acceleration process:} In this case, the position $x(t)$ of the particle evolves via
    \begin{align}
    \frac{dx}{d \tau} = v,~~~~\frac{dv}{d \tau} = \eta (\tau),
    \end{align}
    so that the process becomes non-Markov in $x$-variable. 
    \item \textbf{Ornstein-Uhlenbeck process:} Here, a diffusing particle is placed in a harmonic trap with potential strength $\frac{1}{2}x^2$ so that position of the particle evolves as
    \begin{align}
    \frac{dx}{d \tau} = -x + \eta (\tau).
    \end{align}
    This is also another canonical example of a Markov process.
\end{enumerate}

\end{widetext}

\end{document}